\documentclass[
preprintnumbers,superscriptaddress,nofootinbib,prx]{revtex4-2}
\usepackage{amsfonts,amsmath, amssymb,bm}
\usepackage[utf8]{inputenc}
\usepackage{hyperref}
\usepackage{qcircuit}
\usepackage{enumitem}
\usepackage{graphicx}

\graphicspath{{./figures/}}

\def\bra#1{\left\langle #1\right|}
\def\ket#1{\left| #1\right\rangle}

\newcommand\mytop[2]{\genfrac{}{}{0pt}{}{#1}{#2}}

\usepackage{pgf}
\usepackage{xparse}
\usepackage{tikz}
\usetikzlibrary{arrows,arrows.meta,backgrounds,calc,cd,matrix,math,positioning}

\begin{document}

\preprint{UT-Komaba-22-2}
\date{\today}
\title{Conserved charges in the quantum simulation of integrable spin chains}

\author{Kazunobu Maruyoshi}\email{maruyoshi@st.seikei.ac.jp}
\affiliation{{\it Faculty of Science and Technology, Seikei University, Musashino-shi, Tokyo 180-8633 Japan}}

\author{Takuya Okuda}\email{takuya@hep1.c.u-tokyo.ac.jp}
\affiliation{{\it Graduate School of Arts and Sciences, University of Tokyo,
Komaba, Meguro-ku, Tokyo 153-8902, Japan}}

\author{Juan W. Pedersen}\email{pedersen@hep1.c.u-tokyo.ac.jp}
\affiliation{{\it Graduate School of Arts and Sciences, University of Tokyo,
Komaba, Meguro-ku, Tokyo 153-8902, Japan}}

\author{Ryo Suzuki}\email{rsuzuki.mp@gmail.com}
\affiliation{{\it Shing-Tung Yau Center of Southeast University, 
Xuanwu district, Nanjing, Jiangsu, China}}

\author{Masahito Yamazaki}\email{masahito.yamazaki@ipmu.jp}
\affiliation{{\it Kavli IPMU (WPI), University of Tokyo, Kashiwa, Chiba 277-8583, Japan}}

\author{Yutaka Yoshida}\email{yutakayy@law.meijigakuin.ac.jp}
\affiliation{{\it Faculty of Law, Meiji Gakuin University, Totsuka-ku, Yokohama, Kanagawa, 244-8539, Japan}}

\begin{abstract}

When simulating the time evolution of quantum many-body systems on a digital quantum computer, one faces the challenges of quantum noise and of the Trotter error due to time discretization.  
For certain spin chains, it is possible to discretize the time evolution preserving integrability, so that an extensive set of conserved charges are exactly conserved after discretization.
In this work we implement, on real quantum computers and on classical simulators, the integrable Trotterization of the spin-$1/2$ Heisenberg XXX spin chain.  
We study how quantum noise affects the time evolution of several conserved charges, and observe the decay of the expectation values.
We in addition study the early time behaviors of time evolution, which can potentially be used to benchmark quantum devices and algorithms in the future.  
We also provide an efficient method to generate the conserved charges at higher orders.
\footnote{The author names are listed alphabetically.}

\end{abstract}

\maketitle

\tableofcontents

\section{Introduction}

One of the most promising applications of quantum computing is the simulation of quantum-mechanical systems in Nature with many degrees of freedom \cite{Feynman}. Future quantum computers are expected to provide powerful resources for simulations, which cannot be matched by their classical counterparts.
Ultimately one would like to simulate the time evolution of realistic quantum field theories such as the Standard Model of particle physics, quantum chromodynamics and condensed-matter systems.
At present, however, noise is one of the most significant obstacles in quantum computation.
It is important to understand and reduce its effects on the digital quantum simulation of quantum many-body systems.

When simulating the time evolution of a quantum mechanical system with Hamiltonian $H$ on a digital quantum computer, one often discretizes time $t$ and approximates $e^{-itH}$ by $U^d$, namely $d$ repeated applications of a product of unitary operators $U$ composed of elementary gates~\cite{Lloyd1996UniversalQS}. 
This approximation, called the Lie-Trotter-Suzuki decomposition or Trotterization, introduces a Trotter error.
For given $t$ one can make the Trotter error small by taking the depth $d$ large. 
However, increasing $d$ introduces more errors due to quantum noise. Thus, there is a competition between noise and the Trotter error.

Integrable spin chains can be Trotterized in a way that preserves integrability, called integrable Trotterization \cite{2018PhRvL.121c0606V,2019PhRvL.122o0605L}.  
In this prescription, the original conserved charges including Hamiltonian are deformed, but remain mutually commuting.
Therefore the new conserved charges are free of the Trotter error, and exactly conserved in the discrete time evolution.

The study of integrable spin chains is important in its own right.  They are among the most well-studied quantum many-body systems where many exact results are known, and often serve as useful stepping stones for studying more general non-integrable models.  Integrable spin chains may also be regarded as simple lattice quantum field theories in 1+1 spacetime dimensions, where both time and space are discrete.

In this work, we implement the integrable Trotterization of the spin-$1/2$ Heisenberg XXX spin chain \cite{2018PhRvL.121c0606V} with periodic boundary conditions, on real quantum computers and on classical simulators.  We devise protocols to simulate the discrete time evolution and measure the expectation values of the conserved charges.  
While the Trotter error is under perfect control in the sense that the deformed charges are exactly conserved, we find that the effects of quantum noise are significant, forcing the charges to decay.  We study the behaviors of the decay using classical simulators and the theory of quantum error channels.  

We also suggest that the early-time behaviors of the charges may be used to benchmark quantum devices and algorithms in the future when the noise exists but has smaller effects.  Compared with other benchmarking protocols such as randomized benchmarking~\cite{2005JOptB...7S.347E,2007Sci...317.1893E,PhysRevA.77.012307,PhysRevLett.106.180504,2019PhRvA.100c2328C}, the (non-)conservation of charges will provide intuitive benchmarking measures in the setting directly related to the simulation of quantum field theories.

As a by-product, we obtain explicit expressions for low-degree higher conserved charges of the integrably-Trotterized XXX spin chain. 
It is known that these conserved charges can be recursively generated by the so-called boost relation. Since the conserved charges become quickly complicated as the degree increases, we invent an efficient way to solve the recursion relations and implement it as a classical computer (Mathematica) program.

The simulation of integrable spin chains on digital quantum computers has been intensively studied these days \cite{Cervia:2020fkk,Robbins_2021,Nepomechie_2020,2022JPhA...55e5301V,2022arXiv220103021L,2021PRXQ....2d0329V,2022arXiv220204673S}, and even Lindblad superoperators can be integrable \cite{Ziolkowska:2019ktu,deLeeuw:2021cuk}.\footnote{For a review of quantum simulation for high energy physics, see~\cite{Bauer:2022hpo}.
For a sample of works on the digital simulation in related fields, see~\cite{2019npjQI...5..106S,2020arXiv201007965A,Liu:2021onc,Yu:2022ivm}.
}
To the best of our knowledge our work is the first study of integrable Trotterization on a real quantum device.  
The rest of this paper is organized as follows.
In section \ref{sec:charges} we review the Trotterized XXX spin chain and 
discuss recursion related to the conserved charges.
In section \ref{sec:circuits} we propose quantum circuits for measuring the 
expectation values of the conserved charges.
In section \ref{sec:results} we present the results of our simulations
(both on simulators and on a real device of IBM Quantum).
In section \ref{sec:analysis} we comment on the implications of the simulation results.
We conclude in section \ref{sec:discussion} with a summary and discussions.
We also include several appendices for technical materials including data taken on a large IBM device and on a trapped ion device of IonQ.

\section{Conserved charges in Trotterized XXX spin chains}\label{sec:charges}

\subsection{Integrable Trotterization of the XXX spin chain}
\label{sec:staggered-XXX}

As a prototypical example of an integrable spin chain, we consider
the standard spin-$1/2$ XXX spin chain defined by the Hamiltonian
\begin{equation}
    H = \frac{J}{2} \sum_{j=1}^N \bm{\sigma}_j\cdot \bm{\sigma}_{j+1}
= \frac{J}{2} \sum_{j=1}^N (
    X_jX_{j+1} +  Y_jY_{j+1} +  Z_jZ_{j+1}
    )\,,
\label{def:XXX Hamiltonian}
\end{equation}
where $\bm{\sigma}_j = (X_j,Y_j,Z_j)$ denotes the vector of Pauli operators $(\sigma_x,\sigma_y,\sigma_z)$ on site $j$.
We take $N$ to be even and impose the periodic boundary condition $\bm{\sigma}_{N+1}=\bm{\sigma}_1$.
This model can be exactly solved by the algebraic Bethe ansatz, which can be formulated in terms of the R-matrix and the transfer matrix (See \cite{Faddeev:1996iy,Slavnov:2018kfx} for reviews).

For quantum simulations, we need to discretize the time evolution operator $\exp(-i H t)$ into a sequence of discrete time evolutions.
In~\cite{2018PhRvL.121c0606V}, a deformation of the XXX model was introduced, which we call the Trotterized XXX model. The Trotterized XXX model is a special case of the inhomogeneous XXX model, 
where the inhomogeneities are controlled by the deformation parameter $\delta$.
In this model, we can discretize the time evolution while preserving integrability.

Let us write the $R$-matrix of the original XXX model as $R_{ij}(\lambda) = P_{ij} \check R_{ij}(\lambda)$, where
\begin{equation}\label{eq:def_R_hat}
\check R_{ij}(\lambda) = \frac{1+i\lambda P_{ij}}{1+i\lambda} 
\end{equation}
and $P_{ij}=(1+\bm{\sigma}_i\cdot\bm{\sigma}_j)/2$ acts as a permutation of the qubits at sites $i$ and $j$.
We introduced a formal variable $\lambda$, called the spectral parameter.
The discrete time evolution is generated by $d$ ($\sim$depth) repeated actions of the unitary operator
\begin{equation}\label{eq:evolution-U}
\mathcal{U}(\delta) = 
\left(\prod_{j=1}^{N/2} \check{R}_{2j-1,2j}(\delta)\right)
\left(\prod_{j=1}^{N/2} \check{R}_{2j,2j+1}(\delta)\right) \,,
\end{equation}
where the deformation parameter $\delta$ is regarded as a Trotterized time step,
so that the total time for the evolution is $t=-(\delta/J)d$.
Note that the ordering matters in \eqref{eq:evolution-U}:
the $\check{R}$-matrices within each bracket mutually commute, but those from the two different brackets may not commute with each other.
The continuous time evolution of the original XXX model can be studied by the relation
 \begin{align}
\mathcal{U}(-J t/d)^d = e^{ - it ( H - \frac12 NJ ) } 
\Bigl( 1 + \mathcal{O}(J^2 t^2/d)\Bigr), 
\end{align}
for large $d$ and fixed $J, t$. The second term represents the Trotter error.

Below we take $\delta$ as a real-valued independent parameter.\footnote{The parameter $J$ shows up only in the combination $Jt$, which can always be replaced by $\delta$.}
This unitary operator can be decomposed as\footnote{This relation was first found in the light-cone discretization of integrable QFT \cite{Destri:1987ug}. Recently it has been called integrable Floquet dynamics \cite{Gritsev:2017zdm,Miao:2022dau} and integrable Trotterization \cite{2018PhRvL.121c0606V}.} 
\begin{align}\label{U_decompose}
\mathcal{U}(\delta)=T(-\delta/2)^{-1}T(\delta/2) \,,
\end{align}
where the transfer matrix is defined by 
\begin{equation}\label{def:transfer w. delta}
    T(\lambda)= {\rm tr}_0\left(
    \prod_{1\leq j\leq N}^{\longleftarrow}
    R_{0j}\left(\lambda - (-1)^j \frac{\delta}{2}\right)
    \right) \,.
\end{equation}
Here the product is taken in ascending order from right to left, and the trace is taken in an auxiliary space (ancilla).
The Trotterized XXX model is integrable because the transfer matrices with arbitrary and different values of the spectral parameter mutually commute: $[T(\lambda),T(\mu)]=0$.

\subsection{Recursion relations for conserved charges}
\label{subsec:charges}

The mutual commutativity of the family of transfer matrices~$T(\lambda)$ implies conservation of the charges under the Trotterized time evolution: 
\begin{equation}\label{eq:charges-def}
Q^\pm_n \sim \frac{d^n}{d\lambda^n}\log T(\lambda)\Big|_{\lambda=\pm \delta/2} \,, \qquad (N > 2n+1).
\end{equation}
Throughout the paper, the length $N$ of the spin chain must be greater than $2n+1$ when measuring $Q^\pm_n$, because the conserved charges above this bound are not generated by the simple recursion relation given below.

Let us introduce the precise expressions of the charges $Q^\pm_n$\,, which will be used in the quantum simulation.
For our purposes it is useful to start with the explicit formulas for $Q^\pm_1$ and $Q^\pm_2$ as follows. 
We first define the charges in terms of the densities: 
\begin{eqnarray}
    Q_n^+(\delta) 
    &=&  \sum_{j=1}^{N/2} 
    q^{[n,+]}_{2j-2, 2j-1,\ldots,2j+2n-2}(\delta),
    \label{eq:Qpm} \\
    Q_n^-(\delta) 
    &=&   \sum_{j=1}^{N/2}
    q^{[n,-]}_{2j-1,2j,\ldots,2j+2n-1}(\delta),
    \label{eq:Qpm-qpm}
\end{eqnarray}
which contain Pauli matrices acting on at most $2n+1$ sites.
As shown in~\cite{2018PhRvL.121c0606V},  $q^{[1, \pm]}_{1,2,3}$ and $q^{[2, \pm]}_{1,2,3,4,5}$ are given by
\begin{equation}\label{def:q1 density}
    q^{[1, \pm]}_{1,2,3}(\delta)
     =     {\boldsymbol\sigma}_1\cdot{\boldsymbol\sigma}_2 + {\boldsymbol\sigma}_2\cdot{\boldsymbol\sigma}_3 \mp \delta {\boldsymbol\sigma}_1\cdot ({\boldsymbol\sigma}_2 \times {\boldsymbol\sigma}_3) + \delta^2 {\boldsymbol\sigma}_2 \cdot {\boldsymbol\sigma}_3 \,,
\end{equation}
and 
\begin{eqnarray}\label{def:q2 density}
    q^{[2, \pm]}_{1,2,3,4,5}(\delta)
     &=&   \mp 2 \delta ( {\boldsymbol\sigma}_3\cdot{\boldsymbol\sigma}_4 + {\boldsymbol\sigma}_4\cdot{\boldsymbol\sigma}_5 - {\boldsymbol\sigma}_3\cdot{\boldsymbol\sigma}_5)
     - (1 - \delta^2) {\boldsymbol\sigma}_3\cdot ({\boldsymbol\sigma}_4 \times {\boldsymbol\sigma}_5) - {\boldsymbol\sigma}_2\cdot ({\boldsymbol\sigma}_3 \times {\boldsymbol\sigma}_4) - \delta^2 {\boldsymbol\sigma}_2\cdot ({\boldsymbol\sigma}_3 \times {\boldsymbol\sigma}_5)  
     \nonumber \\
    & & - \delta^2 {\boldsymbol\sigma}_1\cdot ({\boldsymbol\sigma}_3 \times {\boldsymbol\sigma}_4) - \delta^4 {\boldsymbol\sigma}_1\cdot ({\boldsymbol\sigma}_3 \times {\boldsymbol\sigma}_5)  \pm \delta {\boldsymbol\sigma}_2\cdot ({\boldsymbol\sigma}_3 \times {\boldsymbol\sigma}_4 \times {\boldsymbol\sigma}_5)  \pm \delta {\boldsymbol\sigma}_1\cdot ({\boldsymbol\sigma}_2 \times {\boldsymbol\sigma}_3 \times {\boldsymbol\sigma}_4) 
    \nonumber \\
    & & \pm \delta^3 {\boldsymbol\sigma}_1\cdot ({\boldsymbol\sigma}_3 \times {\boldsymbol\sigma}_4 \times {\boldsymbol\sigma}_5) \pm \delta^3 {\boldsymbol\sigma}_1\cdot ({\boldsymbol\sigma}_2 \times {\boldsymbol\sigma}_3 \times {\boldsymbol\sigma}_5) - \delta^2 {\boldsymbol\sigma}_1\cdot ({\boldsymbol\sigma}_2 \times {\boldsymbol\sigma}_3 \times {\boldsymbol\sigma}_4\times {\boldsymbol\sigma}_5) .
\end{eqnarray}
where ${\boldsymbol\sigma}_1\cdot ({\boldsymbol\sigma}_2 \times {\boldsymbol\sigma}_3 \times \dots 
\times {\boldsymbol\sigma}_{\ell-1} \times {\boldsymbol\sigma}_\ell) :={\boldsymbol\sigma}_1\cdot ({\boldsymbol\sigma}_2 \times( {\boldsymbol\sigma}_3 \times( \dots 
\times ({\boldsymbol\sigma}_{\ell-1} \times {\boldsymbol\sigma}_\ell)\cdots)))$.
One finds that $Q_1^\pm$ reduce to the original XXX Hamiltonian \eqref{def:XXX Hamiltonian} at $\delta=0$. 
As can be seen from these examples, 
the charges and densities are defined so that there is no term proportional to the identity.

We compute the higher charges by the recursion relations
\begin{eqnarray}
Q_{n+1}^{\pm} \equiv [ B, Q_{n}^{\pm} ]  ,
  \label{eq:boost}
\end{eqnarray}
where the degree of the new charge should not exceed the bound \eqref{eq:charges-def}.
The boost operator $B$ is defined by \cite{2018PhRvL.121c0606V}\footnote{These definitions of the boost operator and the charge densities are different from those in \cite{2018PhRvL.121c0606V} by a factor $i/(2(1+\delta^2)^n)$.}
\begin{align}
&B:= \sum_{\ell=1}^{N/2} \ell~ \mathbb{R}'_{2\ell-3,2\ell-2|2\ell-1,2\ell},\\
&\mathbb{R}'_{12|34} 
:=   {\boldsymbol\sigma}_1\cdot{\boldsymbol\sigma}_2+{\boldsymbol\sigma}_3\cdot{\boldsymbol\sigma}_4
+2\,{\boldsymbol\sigma}_2 \cdot{\boldsymbol \sigma}_3 + \delta^2 \,{\boldsymbol \sigma}_2 \cdot{\boldsymbol\sigma}_4+\delta^2\,{\boldsymbol \sigma}_1 \cdot{\boldsymbol \sigma}_3 
 + \delta \,{\boldsymbol \sigma}_1 \cdot \big({\boldsymbol \sigma}_2 \times{\boldsymbol \sigma}_3 \big)
-\delta\,{\boldsymbol\sigma}_2 \cdot \big({\boldsymbol \sigma}_3 \times{\boldsymbol \sigma}_4 \big).
\end{align}
One can check that the boost operation is consistent with \eqref{def:q1 density}, \eqref{def:q2 density}.
We wrote Mathematica programs
to compute the charges $Q^\pm_n$ up to $n=6$.
More details on the recursion relation can be found in Appendix~\ref{sec:recursion}.
For $Q_3^+$ we obtain a 
simple expression (\ref{eq:q3+}) for its density.

We will measure $Q^+_n$ and the difference between $Q^+_n(\delta)$ and $Q^-_n(\delta)$.
Since $Q_n^+(\delta=0)=Q_n^-(\delta=0)$,%
\footnote{%
This follows from~(\ref{eq:charges-def}) and the fact that the overall constant is the same for $Q^+_n$ and $Q^-_n$.
}
we normalize the difference as
\begin{equation}\label{def:Q_dif}
Q_n^\text{dif}(\delta):= \big(Q^+_n(\delta)-Q^-_n(\delta)\big)/\delta ,
\end{equation}
which are also the conserved charges that are polynomials in $\delta$ with integer coefficients.

\section{Quantum circuits for conserved-charge measurements}\label{sec:circuits}

We are now ready to discuss our quantum circuits.
Our quantum circuit is divided into three parts:
initialization, time evolution and measurement:

\begin{equation}\label{eq:total-circuit}
\raisebox{-8mm}{
\includegraphics[width=11cm, trim= 0 0 0 30]{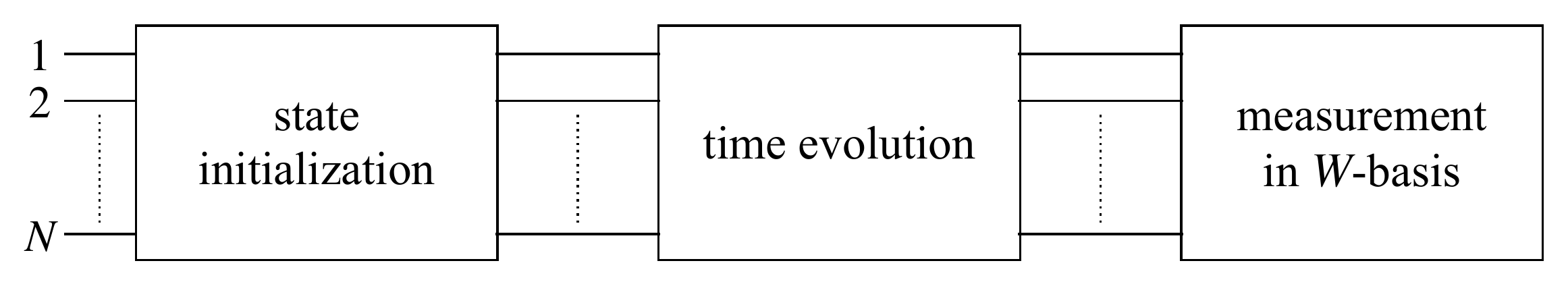}
}
\end{equation}

\subsection{State initialization}
\label{sec:initialization}

The first step is to initialize the state before the dynamical time evolution.
For our discussions of conserved charges, we can choose any state 
as the initial state $|\psi_0\rangle$: the conserved charges will be time-independent under the noiseless time evolution
regardless of the choice of the initial state. 
In our work we consider three types of the initial state:

\begin{itemize}

\item The state $|00 \dots 0\rangle$ in the computational basis.

\item The N\'{e}el state $|010 \dots 01\rangle$.

\item An eigenstate $|s_1 s_2 \dots s_{N} \rangle_{P_1 P_2\dots P_{N}}$ for the operator  ${P_1\otimes P_2\otimes\ldots\otimes P_{N}}$ ($P_j\in\{X,Y,Z\}$) with eigenvalues $(-1)^{s_1+s_2+\dots+s_N}$, with $s_j\in \{0,1\}$.

\end{itemize}

\begin{figure}[t]
\begin{center}
\begin{tabular}{c c}
          
\includegraphics[scale=1]{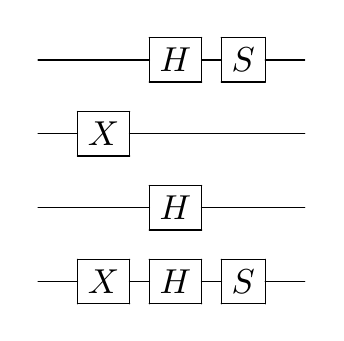}
&
\includegraphics[scale=1]{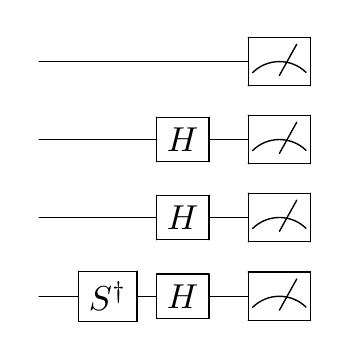}
\\
(a)& 
(b)
\end{tabular}
\caption{(a) State initialization circuit for $|0101\rangle_{YZXY}$.
(b) Measurement circuit for the Pauli word $W=ZXXY$.
}
\label{fig:circuits}
\end{center}
\end{figure}

The third type is the most general, and it includes the first two as special cases.
The state initialization part of the circuit~(\ref{eq:total-circuit})
consists entirely of one-qubit gates, namely  $X_j^{s_j}$ followed by $H_j$ (if $P_j=X_j$), $S_j H_j$ (if $P_j=Y_j$), or none (if $P_j=Z_j$).%
\footnote{In the computational basis, $H=2^{-1/2}\begin{pmatrix} 1&1 \\ 1&-1\end{pmatrix}$, $S={\rm diag}(1,i)$.
Note that $H^{-1}XH=Z$, $(SH)^{-1} Y (SH)=Z$.} 
See FIG.~\ref{fig:circuits}(a).

\subsection{Time evolution}

The second step is to consider the time evolution part of the circuit \eqref{eq:total-circuit}.
The Trotterized time evolution is given by $\mathcal{U}(\delta)$ in (\ref{eq:evolution-U}), which can be expressed as
\tikzset{unitUU/.pic={
\tikzmath{\dex = 0.15; \dey=0.15;} 
\filldraw [thick, fill=gray!10!white, draw=black] (2+\dex, 3-\dey) rectangle (3-\dex, 1+\dey);
\filldraw [thick, fill=gray!10!white, draw=white] (2+\dex, 4+\dey) rectangle (3-\dex, 3+\dey);
\draw [thick]  (2+\dex, 4+\dey) -- (2+\dex, 3+\dey);
\draw [thick]  (3-\dex, 3+\dey) -- (2+\dex, 3+\dey);
\draw [thick]  (3-\dex, 3+\dey) -- (3-\dex, 4+\dey);
\filldraw [thick, fill=gray!10!white, draw=black]  (3+\dex, 4-\dey) rectangle (4-\dex, 2+\dey);
\filldraw [thick, fill=gray!10!white, draw=white] (3+\dex, 2-\dey) rectangle (4-\dex, 1-\dey);
\draw [thick]  (3+\dex, 1-\dey) -- (3+\dex, 2-\dey);
\draw [thick]  (4-\dex, 2-\dey) -- (3+\dex, 2-\dey);
\draw [thick]  (4-\dex, 2-\dey) -- (4-\dex, 1-\dey);
\filldraw [thick, fill=gray!10!white, draw=white] (2+\dex, -1-\dey) rectangle (3-\dex, -2-\dey);
\draw [thick]  (2+\dex, -2-\dey) -- (2+\dex, -1-\dey);
\draw [thick]  (3-\dex, -1-\dey) -- (2+\dex, -1-\dey);
\draw [thick]  (3-\dex, -1-\dey) -- (3-\dex, -2-\dey);
\filldraw [thick, fill=gray!10!white, draw=white] (3+\dex, -1+\dey) rectangle (4-\dex, -2+\dey);
\draw [thick]  (3+\dex, -1+\dey) -- (3+\dex, -2+\dey);
\draw [thick]  (4-\dex, -2+\dey) -- (3+\dex, -2+\dey);
\draw [thick]  (4-\dex, -2+\dey) -- (4-\dex, -1+\dey);
}}
\begin{equation}
\begin{tikzpicture}[baseline=(current bounding box.center), x=4mm, y=4mm]]
\draw [thick] (-3,3.5) -- (5,3.5);
\draw [thick] (-3,2.5) -- (5,2.5);
\draw [thick] (-3,1.5) -- (5,1.5);
\draw [thick] (-3,-1.5) -- (5,-1.5);
\filldraw [thick, fill=white, draw=black]  (-2.5,4) rectangle (4.5,-2);
\node at (1,1) {Time evolution};
\end{tikzpicture}
\quad = \quad
\begin{tikzpicture}[baseline=(current bounding box.center), x=4mm, y=4mm]]
\tikzmath{\dex = 0.15; \dey=0.15; \des=0.2; \dess=0.4;} 
\draw [thick] (-1,3.5) -- (2.4,3.5);
\draw [thick] (-1,2.5) -- (2.4,2.5);
\draw [thick] (-1,1.5) -- (2.4,1.5);
\draw [thick] (3.6,3.5) -- (7,3.5);
\draw [thick] (3.6,2.5) -- (7,2.5);
\draw [thick] (3.6,1.5) -- (7,1.5);
\filldraw [thick, fill=black, draw=black, radius=0.03] (3-\des,3.5) circle;
\filldraw [thick, fill=black, draw=black, radius=0.03] (3,3.5) circle;
\filldraw [thick, fill=black, draw=black, radius=0.03] (3+\des,3.5) circle;
\filldraw [thick, fill=black, draw=black, radius=0.03] (3-\des,2.5) circle;
\filldraw [thick, fill=black, draw=black, radius=0.03] (3,2.5) circle;
\filldraw [thick, fill=black, draw=black, radius=0.03] (3+\des,2.5) circle;
\filldraw [thick, fill=black, draw=black, radius=0.03] (3-\des,1.5) circle;
\filldraw [thick, fill=black, draw=black, radius=0.03] (3,1.5) circle;
\filldraw [thick, fill=black, draw=black, radius=0.03] (3+\des,1.5) circle;
\draw [thick] (-1,-1.5) -- (2.4,-1.5);
\draw [thick] (3.6,-1.5) -- (7,-1.5);
\filldraw [thick, fill=black, draw=black, radius=0.03] (3-\des,-1.5) circle;
\filldraw [thick, fill=black, draw=black, radius=0.03] (3,-1.5) circle;
\filldraw [thick, fill=black, draw=black, radius=0.03] (3+\des,-1.5) circle;
\path pic (a) at (2.2,0) {unitUU};
\filldraw [thick, fill=black, draw=black, radius=0.03] (5, \dess) circle;
\filldraw [thick, fill=black, draw=black, radius=0.03] (5,0) circle;
\filldraw [thick, fill=black, draw=black, radius=0.03] (5,-\dess) circle;
\path pic (a) at (-2.2,0) {unitUU};
\filldraw [thick, fill=black, draw=black, radius=0.03] (1,\dess) circle;
\filldraw [thick, fill=black, draw=black, radius=0.03] (1,0) circle;
\filldraw [thick, fill=black, draw=black, radius=0.03] (1,-\dess) circle;
\end{tikzpicture}
\label{fig:time-evolve U}
\end{equation}
where a rectangle represents an $R$-matrix.
We split the $R$-matrix \eqref{eq:def_R_hat} into a phase and two unitary operators:
\begin{equation}
\check R_{j,j+1}(\delta) = e^{-i\frac{\alpha}{2}} \, e^{i \frac{\alpha}{2} (X_j X_{j+1} + Y_j Y_{j+1})} e^{i \frac{\alpha}{2} Z_j Z_{j+1}} \,, \qquad \delta =\tan\alpha.
\label{eq:R-check-alpha}
\end{equation}
We will ignore the overall phase $e^{-i\frac{\alpha}{2}}$ because it does not affect measurements.\footnote{Usually the overall scalar factor of the R-matrix is neglected in quantum circuits. Thus, our simulation corresponds to any of the integrable spin chains that have the same R-matrix as the Heisenberg XXX spin chain up to a scalar factor. For example, the bi-local deformation modifies the scalar factor, which redefines the conserved charges \cite{Bargheer:2008jt}.}
The two unitary operators in \eqref{eq:R-check-alpha} can be realized in terms of elementary gates as follows:
\begin{align}
\Qcircuit @C=1em @R=1.4em {
& \multigate{1}{
e^{i \frac{\alpha}{2} (X \otimes X + Y\otimes Y)}
} & \qw \\ 
& \ghost{e^{i \frac{\alpha}{2} (X \otimes X + Y\otimes Y)} } & \qw
} \quad &\raisebox{-4mm}{$=$} \quad 
\Qcircuit @C=1em @R=.7em {
& \ctrl{1} & \gate{H} & \ctrl{1} & \gate{R_Z (-\alpha)} & \ctrl{1} & \gate{H}  & \ctrl{1}
& \qw\\
& \targ & \qw & \targ & \gate{R_Z (\alpha)} & \targ & \qw & \targ
& \qw 
} \label{circ:XXYY} \\[3mm]
\Qcircuit @C=1em @R=.8em {
& \multigate{1}{
e^{i \frac{\alpha}{2} Z\otimes Z}
} & \qw \\ 
& \ghost{
e^{i \frac{\alpha}{2} Z\otimes Z}
} & \qw
} \quad &\raisebox{-4mm}{$=$} \quad 
\Qcircuit @C=1em @R=.7em {
& \ctrl{1} & \qw & \ctrl{1} & \qw\\
& \targ & \gate{R_Z (-\alpha)} & \targ & \qw 
}
\label{circ:ZZ}
\end{align}
The rotation gate is defined by $R_Z (\alpha) = e^{-i\alpha Z/2}$, in agreement with {\tt RZGate} in Qiskit.
The time evolution part of the circuit~(\ref{eq:total-circuit}) consists of $d$ repetitions of~(\ref{circ:XXYY}) and~(\ref{circ:ZZ}).
We note that pairs of CNOT gates cancel when 
(\ref{circ:XXYY}) and~(\ref{circ:ZZ}) are combined.\footnote{In fact, three CNOT gates are sufficient to construct this R-matrix as a quantum gate \cite{PhysRevA.69.032315}.}

\subsection{Measurement protocol via Pauli words}\label{sec:measure-Pauli}

Let us next discuss the measurement of conserved charges.
Suppose that we have a quantum state described by a density matrix $\rho$.
We then wish to evaluate the expectation value $\langle Q\rangle = {\rm tr}( \rho Q)$ of the charge $Q$,
as well as its statistical uncertainty. 

We review below a measurement protocol via ``Pauli words''.
See, for example, \cite{2020PhRvX..10c1064B}.
Recall that the charge density \eqref{eq:Qpm}, \eqref{eq:Qpm-qpm} is a sum over the product of Pauli matrices, $P, P', \dots$. Our idea is to measure another Pauli operator (Pauli word) $W$ which contains $P, P'$ as a subset. The measurement of $W$ tells us the expectation values of $P,P', \dots$ all at once.
To compute the charge density, we need to choose an appropriate set of Pauli words and repeatedly run the circuit for all the Pauli words in the set.

According to the recursion relation \eqref{eq:boost} a general conserved charge $Q$ can be expanded as
\begin{equation}\label{eq:charge-Pauli}
Q = \sum_{P\in \{I,X,Y,Z\}^{\otimes N}} c_{Q,P}P \,, \quad c_{Q,P} \in \mathbb{C}\,.
\end{equation}
The problem thus reduces to the evaluation of the expectation values~$\langle P\rangle={\rm tr}(\rho P)$ for all the Pauli operators $P$ that appear in the expansion~(\ref{eq:charge-Pauli}) of a given conserved charge~$Q$.

Let us consider a word $W=L_1\ldots L_N$ whose letters $L_j$ ($j=1,\ldots,N$) are one of  $X$, $Y$, and $Z$ (but not $I$). 
We use such a ``Pauli word" to specify a measurement basis ($W$-basis) which diagonalizes the product operator $L_1\otimes\ldots\otimes L_N$~\cite{2020PhRvX..10c1064B}.
For each qubit, the letter $Z$ corresponds to the computational basis $\{|0\rangle, |1\rangle\}$, $X$  to $\{|+\rangle =H |0\rangle,|-\rangle = H |1\rangle\}$, and $Y$ to $\{SH |0\rangle, SH |1\rangle\}$, with the eigenvalues $\{+1,-1\}$ respectively.
We say that a Pauli operator $P=P_1\otimes\ldots\otimes P_N\in  \{I,X,Y,Z\}^{\otimes N}$
 is contained in the Pauli word $W$ (and write $W\supset P$) if $P_j= L_j$ for all $j$ such that $P_j\neq I$.%
\footnote{%
For example, both $ X\otimes I\otimes Z\otimes I =  X_1 Z_3$ and $X\otimes X\otimes I\otimes I= X_1 X_2$ are contained in $XXZY$.}
For given~$Q$, we need to find a set $\mathcal{S}$ of Pauli words~$W$ such that all the Pauli operators $P$ that appear in~(\ref{eq:charge-Pauli}) are contained in some $W\in\mathcal{S}$.  Given such $\mathcal{S}$, measurements along the bases specified by all words in $\mathcal{S}$ are sufficient to evaluate~$\langle Q\rangle$.
Note that we allow each $P$ to be contained in more than one word.
To generate an appropriate $\mathcal{S}$, we make use of a program (\verb9data_acquisition_shadow.py9) provided by H.-Y.~Huang~\cite{data_acquisition_shadow}.%
\footnote{%
The code was developed to implement classical shadow~\cite{2020NatPh..16.1050H} and derandomization~\cite{2021PhRvL.127c0503H} schemes. Note that it is not guaranteed that this code generates the most efficient Pauli word set $\mathcal{S}$ for any observables.
}

The standard computational basis measurement can be converted to the $W$-basis measurement by including the corresponding one-qubit gates.
These gates are the inverse of those considered in Section~\ref{sec:initialization}.
For $W=L_1\ldots L_N$ we include $H_j$ if $L_j=X_j$, $H_j S_j^\dagger$ if $L_j=Y_j$, and none if $L_j=Z_j$, for each $j=1,\ldots,N$.
See FIG.~\ref{fig:circuits}(b).

Suppose that we execute the circuit $n_W$ times in the basis specified by $W \in \mathcal{S}$.
In each execution, the measurement outcome $b_j\in\{0,1\}$ for the $j$-th qubit means that the eigenvalue $(-1)^{b_j}$ of the single-qubit Pauli operator $L_j\in\{X,Y,Z\}$ is observed. This defines a bit string $\bm{b}=b_1\ldots b_N\in \{0,1\}^N$.
We perform $n_W$ measurements in the $W$-basis.
The result for the $j$-th qubit in the $i$-th measurement is denoted by $b_{j}^{(W, i)}$. 
As the estimator of~$\langle Q\rangle$ we take
\begin{equation} \label{eq:estimator-Q}
\langle Q\rangle_{\rm est} := 
\sum_{P\in \{I,X,Y,Z\}^{\otimes N}} c_{Q,P} \langle P\rangle_{\rm est} \,,
\end{equation}
where\footnote{%
For any hermitian $\mathcal{O}$, $\langle\mathcal{O}\rangle = {\rm tr}(\rho\mathcal{O})=\sum_a\mathcal{O}_a \langle B_a|\rho|B_a\rangle$, with $|B_a\rangle$ the orthonormalized eigenstates of $\mathcal{O}$ with eigenvalues $\mathcal{O}_a$.
}
\begin{equation}\label{eq:P-estimator}
\langle P\rangle_{\rm est} :=
\frac{1}{n_P} \sum_{ W\supset P} 
\sum_{i=1}^{n_W} \mathop{\prod_{\mytop{j=1}{w_{j}(P) \neq I}}^{N}}
\hspace{-1mm} (-1)^{b_{j}^{(W, i)}} \,.
\end{equation}
Here $n_P:=\sum_{ W\supset P} n_W $ and $w_j(P)=P_j$ for $P=P_1\otimes\ldots\otimes P_N$.

By grouping together $P, P', \dots $ in one word $W$, the measurement results for $P, P', \dots$ are no longer statistically independent. 
In Appendix~\ref{app:stat-unc-est}, we obtain an unbiased estimator of the variance in $\langle Q\rangle_{\rm est}$ by taking this effect into account.
The unbiased estimator is
\begin{equation}\label{eq:sQ2}
  s_{Q}^2
  := \sum_{P,P^\prime}  \frac{n_{P,P^\prime}}{n_{P} n_{P^{\prime}}}\frac{ c_{Q, P} c_{Q, P^{\prime}}}{n_{P,P^\prime} - 1} \sum_{W \supset P, P^{\prime}}\sum_{i=1}^{n_{W}}
  \left(
  \Pi_{P,W,i} -\langle P\rangle_{\mathrm{est}}^{P^{\prime}}\right)\left(
  \Pi_{P',W,i} -\left\langle P^{\prime}\right\rangle_{\mathrm{est}}^{P}\right) \,,
\end{equation}
where $n_{P,P'}:=\sum_{W\supset P,P'}n_W$ and
\begin{equation}
 \Pi_{P,W,i} 
:= \prod_{\mytop{j=1}{w_{j}(P) \neq I}}^{N}(-1)^{b_{j}^{(W, i)}} \,,
\qquad
  \langle P\rangle_{\text {est }}^{P^{\prime}}:=\frac{1}{n_{P,P^\prime}} \sum_{W \supset P, P^{\prime}} 
  \Pi_{P,W,i} \,.
\end{equation}
We use its square-root $s_{Q}$ to estimate the statistical uncertainty.

\section{Simulation results}\label{sec:results}

In this section we present the results of our simulations on a real device as well as on classical simulators.
In the ideal simulation, the conserved charges~$Q_n^+$ and $Q_n^{\rm dif}$ remain constant under time evolution.
On real NISQ devices, however, the charges do not stay constant due to noise.
To analyze such effects, we computed the time evolution of conserved charges and performed quantum state tomography using classical simulators.

We used Qiskit --- IBM's open-source software development kit for quantum computers \cite{Qiskit_GitHub} --- to create circuits and execute jobs on the real quantum device {\tt ibm\_kawasaki} of IBM.
We also used Qiskit to perform classically-emulated quantum computation.
Throughout this section, the error bar represents the statistical uncertainty $s_{Q}$ discussed in Section~\ref{sec:measure-Pauli}, and the value of $\alpha$ in \eqref{eq:R-check-alpha} is chosen to be $0.3$.
In Appendix~\ref{app:IonQ} we give the results of quantum simulations run on the trapped Ion quantum devices of IonQ.

\subsection{Simulation results on a real device}
\label{sec:real-devices}

\begin{figure}[t]
\begin{center}
\begin{tabular}{cc}
 \includegraphics[scale=0.7]{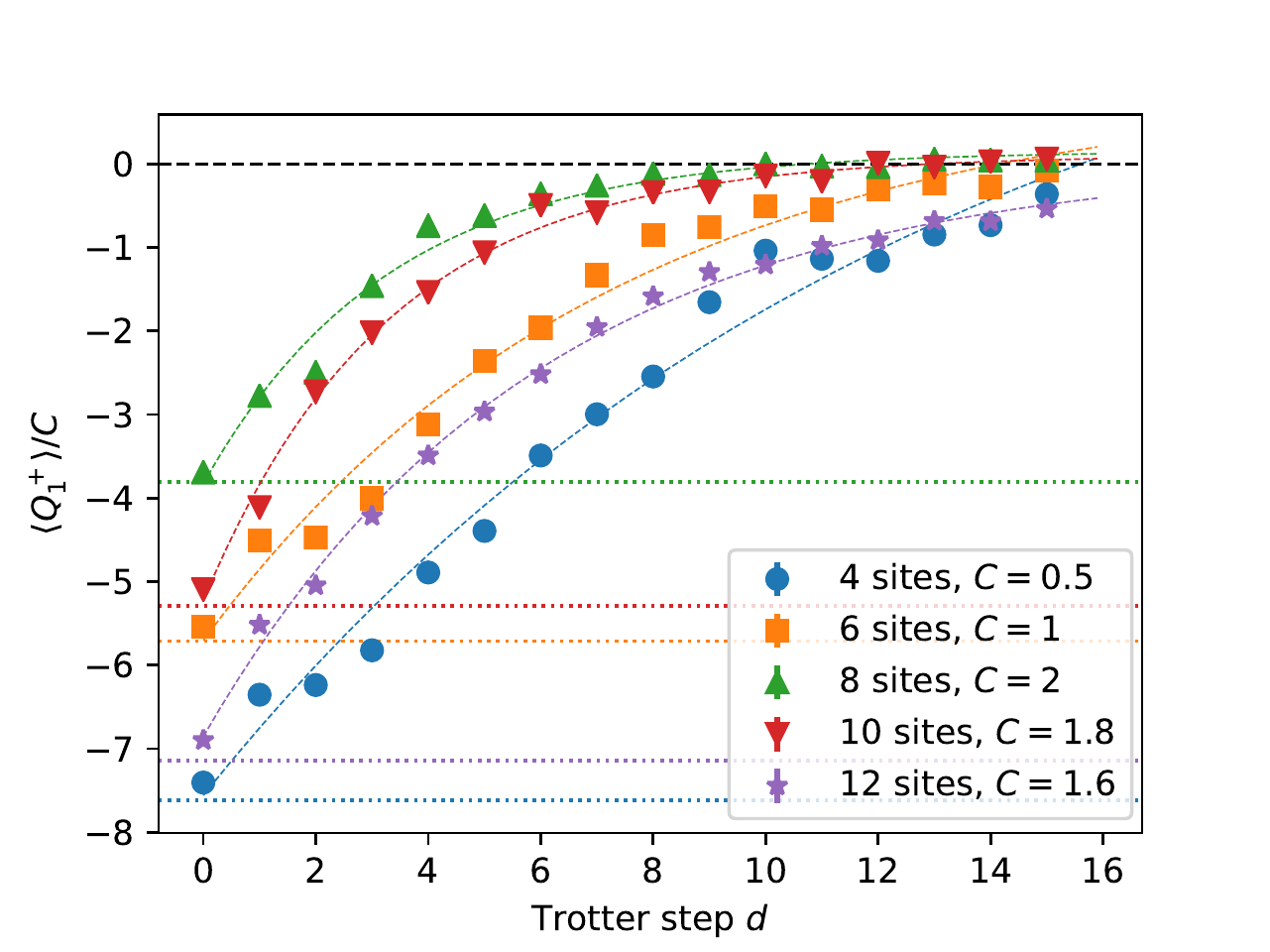}
 &\hspace{-7mm}
 \includegraphics[scale=0.7]{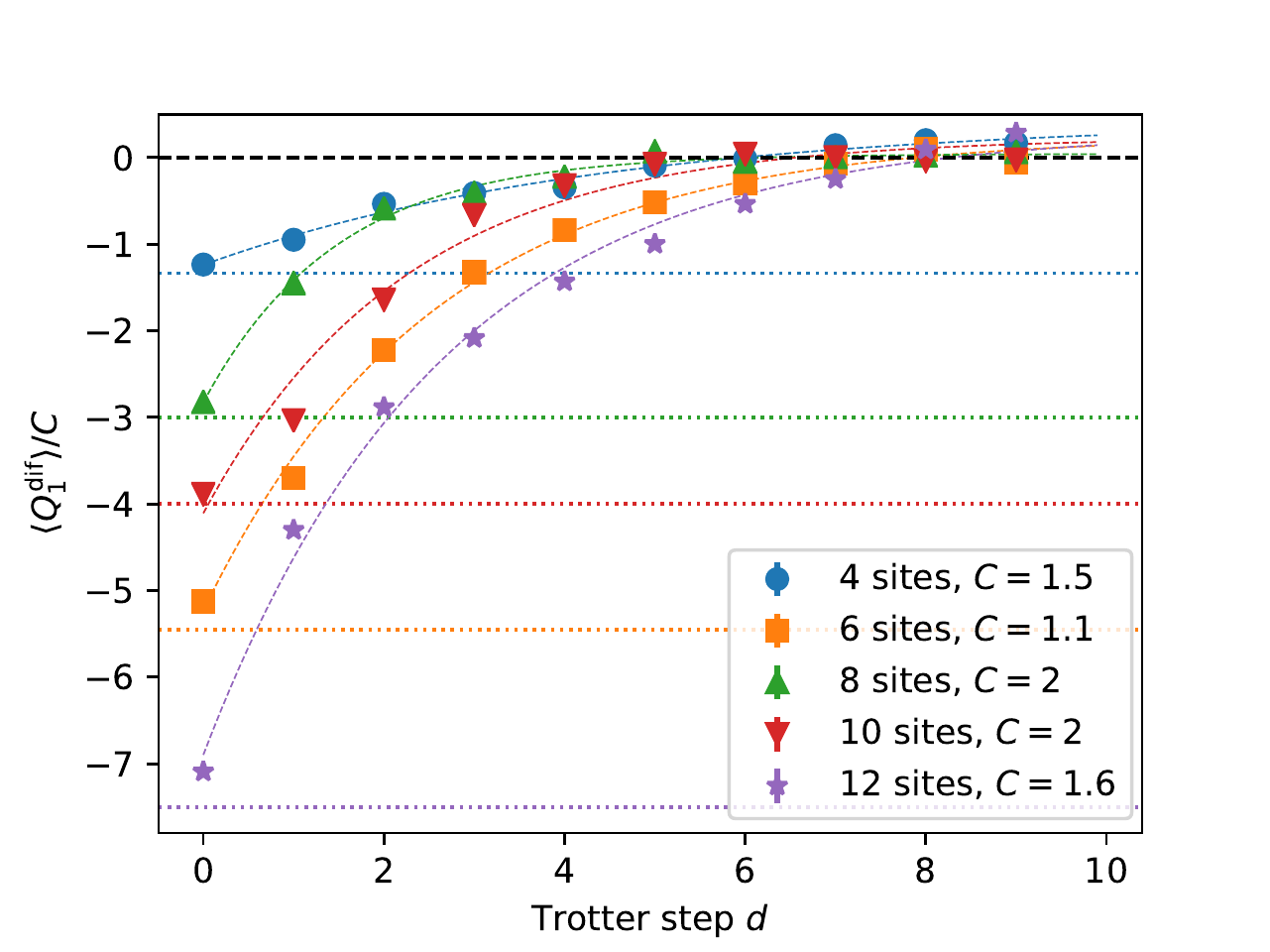}
 \\
(a)&(b)
\end{tabular}

\caption{The {\tt ibm\_kawasaki} simulation results. (a) Results for charge $Q_1^+$ with the initial state $|\text{N\'eel}\rangle =|01\ldots 01\rangle$. 
Each data set is rescaled by a factor $C$, chosen conveniently for better visibility.
The noiseless values of the charges are respectively -3.8, -5.7, -7.6, -9.5, and -11.4 for 4, 6, 8, 10, and 12 sites.
Their rescaled values are indicated by the dotted horizontal lines.
(b) Results for $Q_1^{\rm dif}$ with the following initial states: $|0000\rangle_{YZXX}$ for 4 sites, $|000000\rangle_{YZXYZX}$ for 6 sites, $|00000000\rangle_{YZXYZXYX}$ for 8 sites,  $|0000000000\rangle_{YZXYZXYZXX}$ for 10 sites, and $|000000000000\rangle_{YZXYZXYZXYZX}$ for 12 sites.
The noiseless values of the charges are respectively -2.0, -6.0, -6.0, -8.0, and -12.0 for 4, 6, 8, 10, and 12 sites before rescaling.
In both (a) and (b), error bars representing statistical uncertainties are small and hidden by markers, and the dashed curves indicate the fitted functions $c_1 e^{-\gamma d} + c_2$ with constants $c_1$, $c_2$, and $\gamma$.
The total number of circuit executions (shots) for each value of $d$ was about $10^5$ for all data in (a) and (b).
}
\label{fig:IBM-kawasaki-results}
\end{center}
\end{figure}

We ran simulations on the IBM quantum device {\tt ibm\_kawasaki} with 27 superconducting qubits.
FIG.~\ref{fig:IBM-kawasaki-results} shows the simulation results for charge $Q_1^+$ and $Q_1^\text{dif}$ on $4$, $6$, $8$, $10$, and $12$ sites.%
\footnote{%
See Appendix~\ref{app:IBM} for how the sites of the model are mapped to the physical qubits of the quantum device.
For the simulations on {\tt ibm\_kawasaki}, the physical qubits form a circular loop only for 12 sites. For the other site numbers, the $CX$ gates on non-neighboring qubits are decomposed (transpiled) into a combination of the $CX$ gates on neighboring qubits.
}
The values of the charges exhibit decay as a function of the Trotter step $d$.
While the statistical uncertainties are very small, the computed expected values fluctuate as $t$ varies and do not lie on smooth curves.
This suggests the existence of another type of systematic error.

In Appendix~\ref{app:simulation-details}, we provide further technical details of the simulations, and examine the aforementioned systematic errors.

\subsection{Classically-emulated quantum computation with large noise rates}\label{subsec:large_noise}

In this subsection we discuss the classically emulated quantum computation with quantum errors.
We consider two kinds of noise: 1) depolarizing errors and 2) combined amplitude-and-phase damping errors.
The errors are probabilistically inserted after each gate operation.
Recall that our circuit is mostly composed of the time evolution part, which is a product of $R$-matrices between adjacent qubits as in \eqref{fig:time-evolve U}.
The R-matrix circuit has two kinds of one-qubit gates $H$ and $R_Z(\pm\alpha)$, together with the two-qubit gates $CX$. Thus, the errors quickly and uniformly spread over the entire circuit during the time evolution.

The state preparation and measurement parts also have one-qubit gates ($X$, $H$, and $S^\dagger$). 
The conserved charges $Q_n^+$ and $Q_n^\text{dif}$ defined in (\ref{def:Q_dif}) are measured.

\subsubsection{Simulations with depolarizing errors}
\label{sec:depo}

Let us first consider the noise model with depolarizing errors.
A single-qubit depolarizing channel with a rate $0\leq p\leq 1$ acts on the density matrix $\rho$ as\footnote{%
If ${\rm tr}\rho=1$ we can write $\Phi_\text{depo}(\rho)  = (1-p)\rho + p I/2$.
}
\begin{equation}\label{eq:dep-1q}
\Phi_\text{depo}(\rho) =
\sum_{j=1}^4 D_j \rho D_j^\dagger
\end{equation}
with Kraus operators
\begin{equation}\label{eq:dep_Kraus}
D_1 = \sqrt{1-\frac{3p}{4}} \, I
\,,\quad
D_2 =\sqrt{\frac{p}{4}} \,X
\,,\quad
D_3 =\sqrt{\frac{p}{4}} \,Y
\,,\quad
D_4  =\sqrt{\frac{p}{4}} \,Z
\,.
\end{equation}
In the simulation, after each one-qubit gate, we apply $\Phi_\text{depo}$ with $p=0.0013$ on the qubit.  
After each two-qubit (CNOT) gate, we apply the two-qubit channel $\Phi_\text{depo} \otimes \Phi_\text{depo}$ with $p=0.013$.

\begin{figure}[t]
\begin{center}
\begin{tabular}{c c}
          
\includegraphics[scale=0.7]{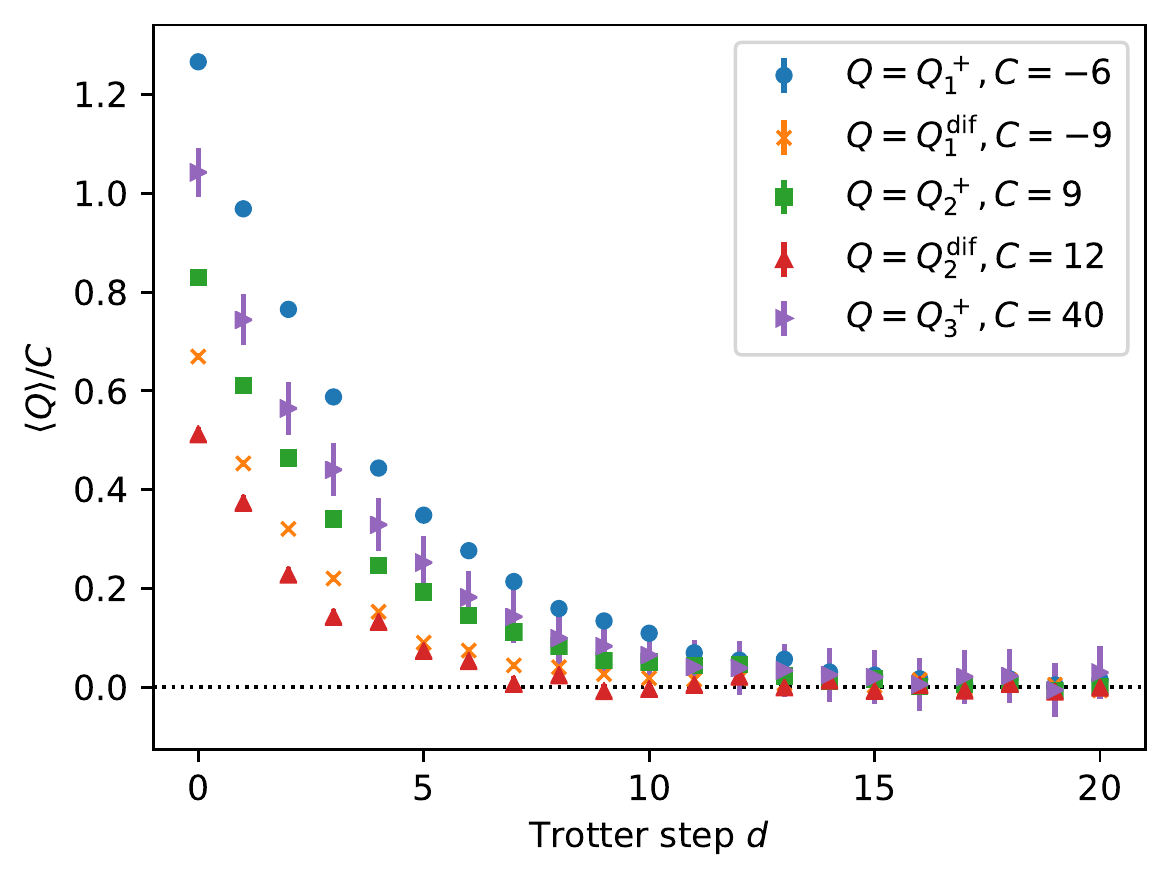}
&
\hspace{3mm}
\includegraphics[scale=0.7]{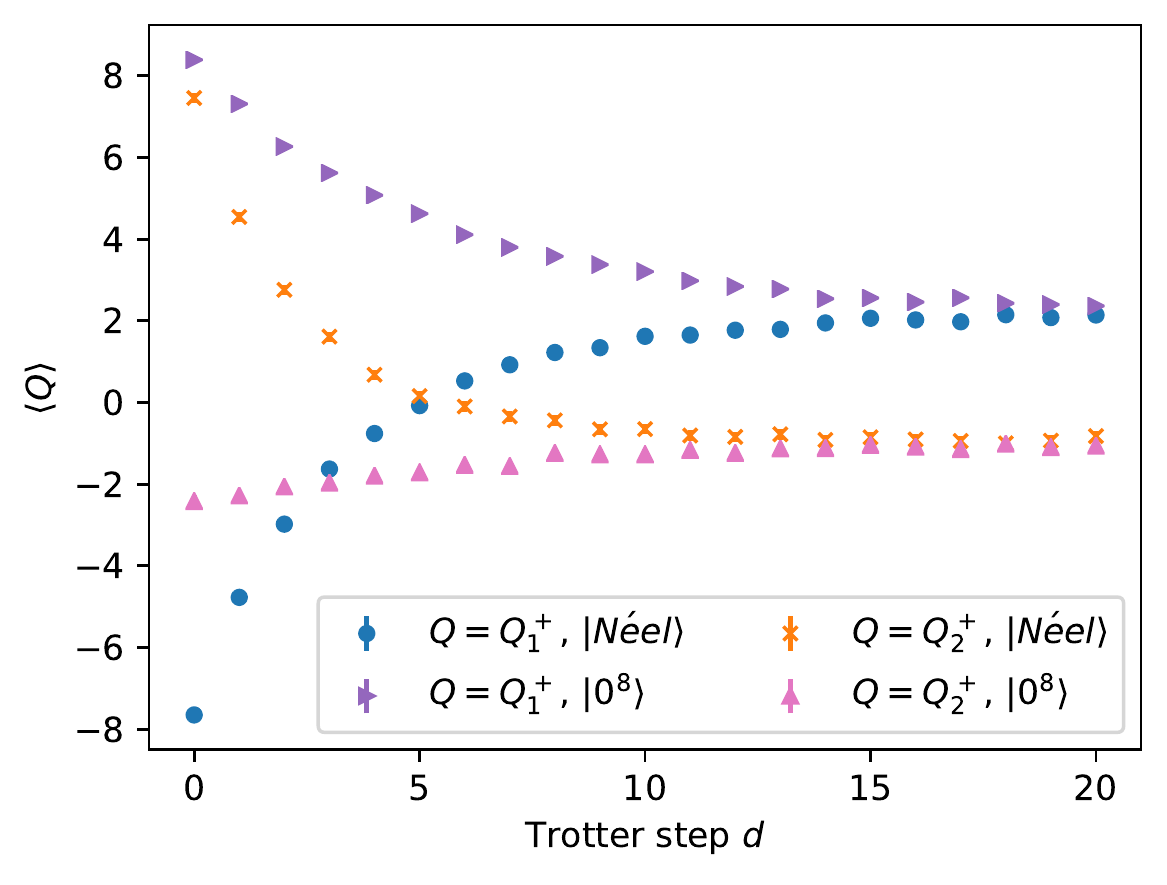}
\\
\hspace{10mm}(a)& 
\hspace{12mm} (b)
\end{tabular}
\caption{(a) Time evolution of conserved charges on 8 sites, under the influence of depolarizing errors.
The values of the rescaling factor $C$ are chosen to display data in the order of decay rates $\gamma$.
The initial state is $|\text{N\'eel}\rangle=|01010101\rangle$ for $Q=Q_n^+$, and $|00000000\rangle_{YZXYZXYX}$ for $Q=Q_n^\text{dif}$.
(b)  Time evolution of conserved charges on 8 sites, under the influence of amplitude and phase damping errors.
The initial states indicated in the legends are $|\text{N\'eel}\rangle=|01010101\rangle$ and $|0^8\rangle=|00000000\rangle$.
Error bars representing statistical uncertainties are small and are hidden by the markers.
}
\label{fig:depo-damp}
\end{center}
\end{figure}

In~FIG.~\ref{fig:depo-damp}(a) we show the results of classically emulated quantum simulations with depolarizing errors.
We see that the expectation value $\langle Q\rangle$ of the conserved charge $Q\in \{Q_1^+,Q_1^\text{dif},Q_2^+,Q_2^\text{dif},Q_3^+\}$ is, as a function of the Trotter step~$d$,
approximated by an exponential $c_1 e^{-\gamma d}$ for some constant $c_1$. 
In particular the expectation value $\langle Q\rangle$ approaches zero asymptotically for large $d$.
This suggests that the final state is the completely mixed state, in which the expectation value $\langle Q\rangle$ of any traceless conserved charge $Q$ vanishes.

\subsubsection{Simulations with amplitude-and-phase damping errors}\label{sec:damp}

We define the combined amplitude-and-phase damping error, with rates $\lambda_a$ and $\lambda_p$, on a single qubit as%
\footnote{%
The error $\Phi_\text{damp}$ is a composition of the amplitude damping error and the phase damping error, which commute.
The rates $\gamma$ and $\lambda$ for the latter two errors in the standard parameterization~\cite{nielsen2002quantum} are related to $\lambda_a$ and $\lambda_p$ as $\gamma=\lambda_a$, $\lambda=\lambda_p(1-\lambda_a)$. 
}
\begin{equation}\label{eq:damp}
\Phi_\text{damp}(\rho) = \sum_{j=1}^3 D_j \rho D_j^\dagger  \,,
\end{equation}
where $\dagger$ denotes hermitian conjugation, and the Kraus operators are given as
\begin{equation}\label{eq:damp_Kraus}
D_1 =
\begin{pmatrix}    
1&0\\
0&\sqrt{1-\lambda_a-\lambda_p}
\end{pmatrix}
\,,\quad
D_2 = 
\begin{pmatrix}    
0&\sqrt{\lambda_a}\\
0&0
\end{pmatrix}
\,,\quad
D_3 = 
\begin{pmatrix}    
0&0\\
0&\sqrt{\lambda_p}
\end{pmatrix}
\,.
\end{equation}
In our noise model we choose the values $\lambda_a=\lambda_p=0.018$.
After each two-qubit gate, we apply $\Phi_\text{damp} \otimes \Phi_\text{damp}$\,.
For simplicity, we let one-qubit gates be noise-free.

In~FIG.~\ref{fig:depo-damp}(b) we show the results of classically emulated quantum simulations with amplitude and phase damping errors.
The expectation value $\langle Q\rangle$ of a conserved charge $Q\in \{Q_1^+,Q_2^+\}$ is approximated by $c_1 e^{-\gamma t}+c_2$ for some constants $c_1$ and $c_2$.
The constant $c_2$ is non-zero for some charges unlike the case of depolarizing errors, and it does not depend on the initial state.
These conclusions also hold for the results for $Q\in\{Q_1^\text{dif},Q_2^\text{dif},Q_3^+\}$, which we omit in FIG.~\ref{fig:depo-damp}(b) to avoid clutter.
We conclude that the final state is not the completely mixed state.

We will strengthen our understanding by quantum state tomography in Section~\ref{sec:tomography} and the  spectrum analysis of the quantum channels in Section~\ref{subsec:early}.

\subsection{Quantum state tomography}\label{sec:tomography}

We have observed that the expectation values of all conserved charges at late times approach a constant value independent of the choice of the initial state. This indicates that the density matrix of the underlying systems gets closer to each other. We perform another noisy simulation for the quantum state tomography (QST), which is the procedure of reconstructing the density matrix from measurements.

\begin{figure}[t]
\begin{center}
\begin{tabular}{c c}          
\includegraphics[scale=0.75]{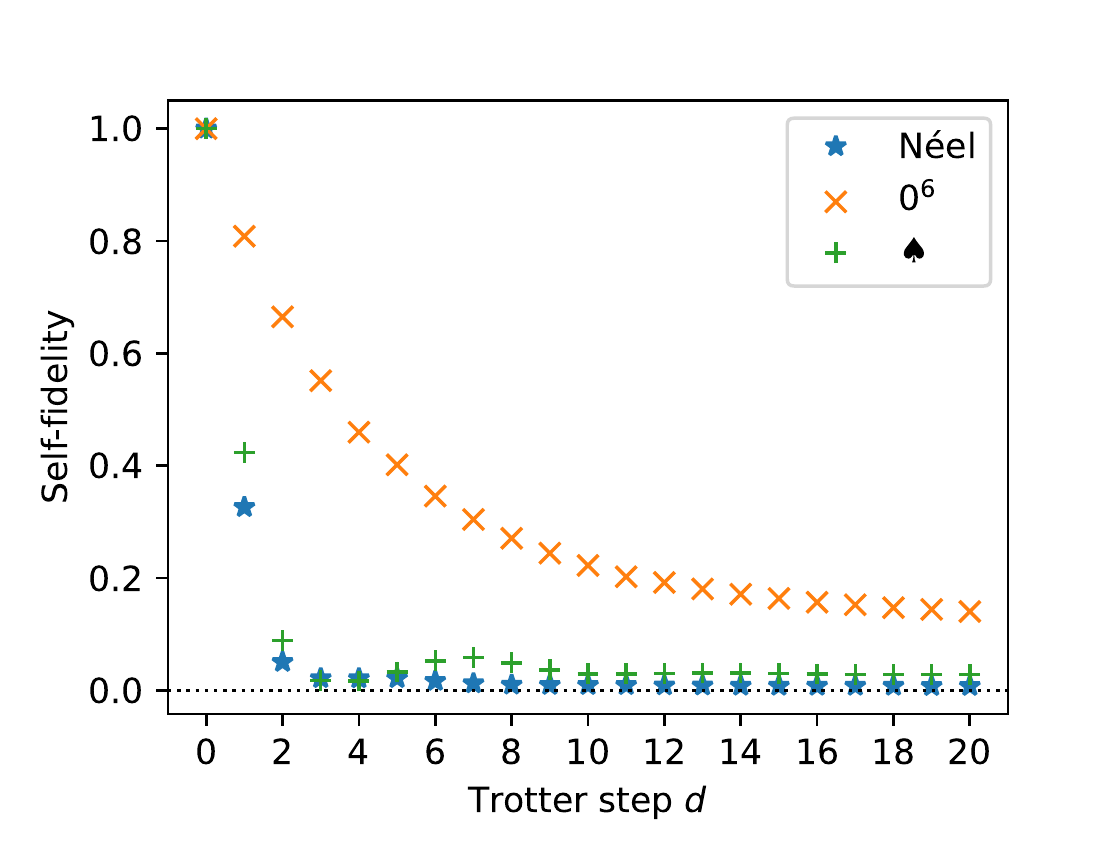}
&
\includegraphics[scale=0.75]{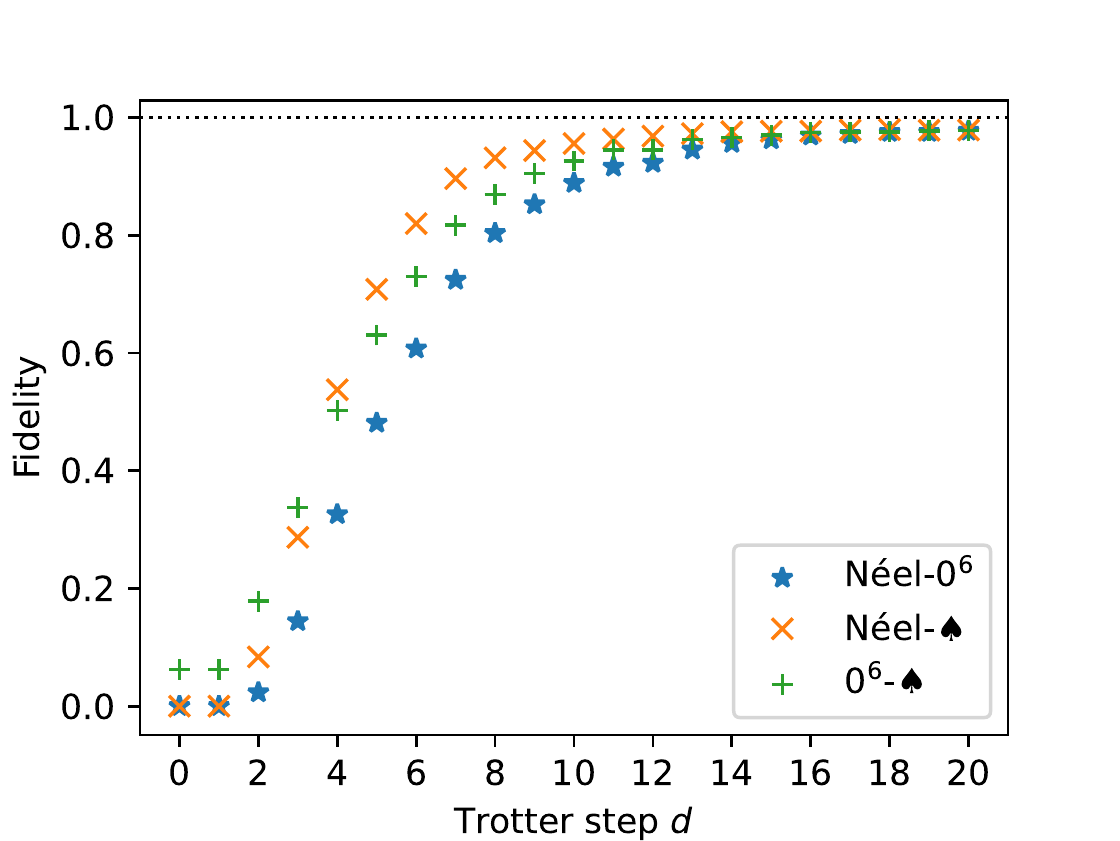}
\\
(a)& 
(b)
\end{tabular}
\caption{(a) Time evolution of self-fidelity on 6 sites, under the influence of amplitude- and phase-damping errors. The error rates are both $0.018$. 
The initial states are $|\text{N\'eel}\rangle=|010101\rangle$, $|0^6\rangle=|000000\rangle$, and $|\spadesuit \rangle=|000000\rangle_{YZXYZX}$\,. We performed 40,000 shots for each measurement observable, and the error bars are negligibly small.
(b)  Time evolution of fidelity between two initial states under the same setup as (a).
The final fidelity values are more than 0.97.
}
\label{fig:tomo}
\end{center}
\end{figure}

In QST, we perform repeated measurements in some basis, and reconstruct the density matrix by solving an optimization problem.
We employed {\tt qiskit-experiments} framework to perform this procedure.\footnote{At the time of writing {\tt qiskit-experiments} framework is under development, and the implementation may change in the future.} 
In particular, we chose the Pauli measurement basis and applied the fitter {\tt fitters.cvxpy\_gaussian\_lstsq} to solve the optimization problem. 
We performed $S=40,000$ shots for each element of the Pauli measurement basis.\footnote{There are two reasons to choose a large number of shots. First, it becomes easier to solve the optimization problem when the statistical errors are smaller. It also saves time, because we can parallelize the measurement of different observables. Second, as the number of sites increases, the final fidelity values in FIG.~\ref{fig:tomo}(b) becomes more sensitive to statistical errors.}
Since QST requires large computational resources, we performed a simulation on the system with 6 sites.
Recently, another efficient method for the full quantum state tomography has been proposed in \cite{Kim_2021}.
See Appendix \ref{app:qst} for the details of the quantum state tomography algorithm.

FIG.~\ref{fig:tomo}(a) shows the self-fidelity for three choices of the initial state, namely the fidelity between the ideal density matrix constructed from the gate set at $d=0$, and the density matrix after Trotter time evolution. 
\begin{equation}
(\text{Self-fidelity})_s (d) = 
F (\rho_s^{\rm ideal} (d=0), \rho_s (d)), \qquad
s \in \{ \text{N\'eel} , 0^6, \spadesuit \}
\end{equation}
where the state fidelity $F$ is defined by $F(\rho,\sigma)= ({\rm tr} \sqrt{ \sqrt{\rho} \sigma \sqrt{\rho} } )^2$.
This result shows that the final state, namely the fixed point of the unitary evolution \eqref{eq:evolution-U} under the influence of the error channel \eqref{eq:damp}, is close but not identical to $|0^6\rangle$.
There are small bumps around $d \sim 7$ in the figure, which seem to be related to our choice of $\alpha=0.3$.

FIG.~\ref{fig:tomo}(b) shows the fidelity between a pair of the density matrices at step $d$, evolved from different initial states,
\begin{equation}
(\text{Fidelity})_{s, s'} (d) = 
F (\rho_s (d), \rho_{s'} (d)), \qquad
s, s' \in \{ \text{N\'eel} , 0^6, \spadesuit \}.
\end{equation}
The fidelity for all pairs becomes close to 1 at a late time, and we expect that the final fidelity approaches 1 as we increase the number of shots.
This fact suggests that all states approach a fixed point of the noisy time-evolution operator.
Clearly, when the density matrices are identical, the expectation values of any observables must also be identical.

\subsection{Classically emulated quantum simulations with small noise rates}
\label{sec:classical-sim-small-noise}

\begin{figure}[t]
\begin{center}
\includegraphics[scale=0.7]{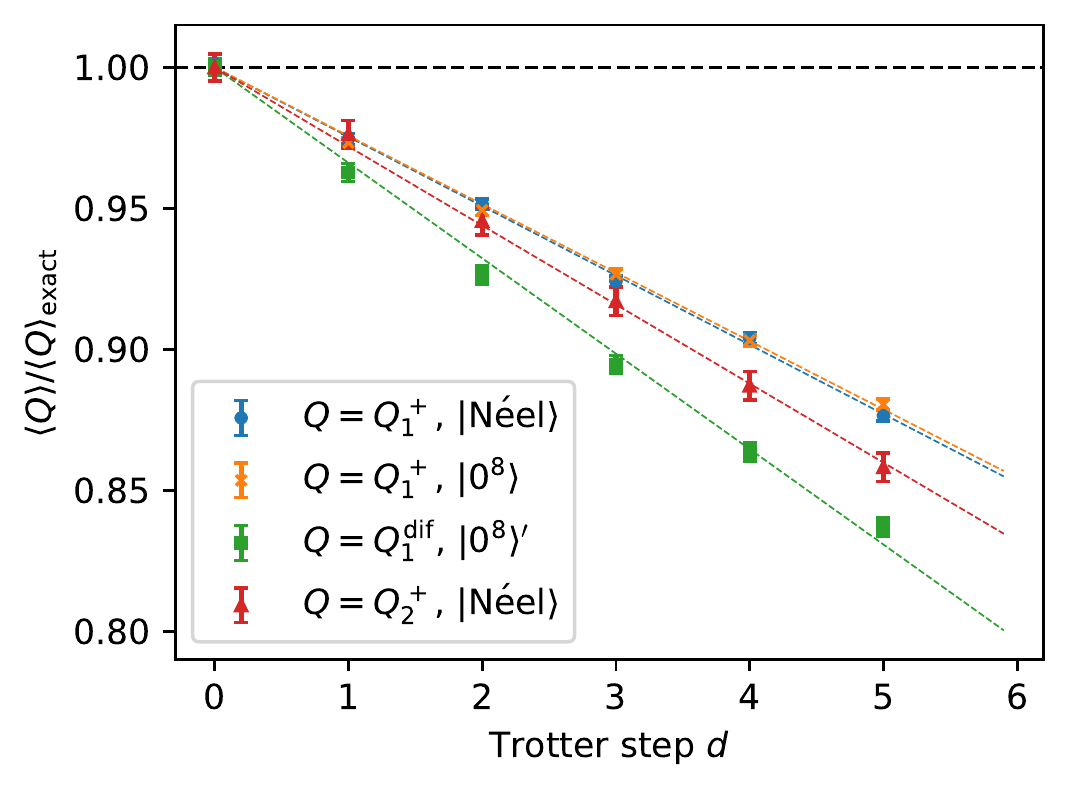}
\caption{Time evolution of conserved charges on 8 sites, under the influence of depolarizing errors, computed by a classical simulator.
The initial states indicated in the legends are $|\text{N\'eel}\rangle=|01010101\rangle$,  $|0^8\rangle = |00000000\rangle$, $|0^8\rangle' = |00000000\rangle_{YZXYZXYX}$.
The error rates are 1/10 of those for FIG.~\ref{fig:depo-damp}(a).
Namely,
after each one-qubit gate we apply $\Phi_\text{depo}$ with $p=0.00013$ on the qubit, and after each two-qubit gate we apply $\Phi_\text{depo}$ with $p= 0.0013$ on each of the two qubits.
We fit the data by dotted straight lines.
}
\label{fig:early-time}
\end{center}
\end{figure}

According to our simulation results on a real device in Section~\ref{sec:real-devices}, the conserved charges rapidly decay in the 
first few Trotter steps. We expect that this initial decay 
will be slower in less noisy devices in the FTQC era,
and deviates from the initial value by small corrections. 
In this case we expect that we can quantify the initial change of the charge $Q$ 
by a linear function for a small number of Trotter steps:
\begin{align}\label{eq:Q_linear_early}
\langle Q \rangle_{d} \sim \langle Q \rangle_{d=0} (1  - \beta   d ) \quad
\quad \textrm{($d$  small)}  \,.
\end{align}
(In practice it is useful to choose the initial state and the charge so that $\langle Q \rangle_{d=0} \ne 0$,
in which case it is easier to extract the linear behavior \eqref{eq:Q_linear_early} 
from numerical results.)

In FIG.~\ref{fig:early-time} we present classical simulation results for small noise rates.
We can conclude that the values of $\beta$ 
are in general different for different choices of charges and for different initial states.
We also find that the values of $\beta$ for $Q_1^{\rm diff}$ and $Q_2^+$ are larger
than those for $Q_1^+$. While data is limited and a more thorough analysis is necessary,
one is tempted to conjecture that one obtains larger values of $\beta$
by considering $Q_n^{\pm}$ for larger values of $n$, and also by 
replacing  $Q_n^{\pm}$ by  $Q_n^{\rm diff}$.
The measurement of $Q_n^{\pm}$ with a larger $n$ can detect more non-local quantum noise.

\section{Analysis of the simulation results}\label{sec:analysis}

We now discuss the implications of the simulation results.
Since the available computational resources are not limitless, some of our results also include significant statistical uncertainties.
Still, our simulation results have indicated some clear patterns already.

In our simulations we have observed two characteristic features at different Trotter steps:

\begin{itemize}
\item ``Late-time behavior'': at large Trotter steps the value of $Q_n^{\pm}$ approaches a constant value.
\item ``Early-time behavior": at small Trotter steps we have a rapid decay of $Q_n^{\pm}$.
\end{itemize}

\subsection{Exponential decays at late times}\label{subsec:early}

In our simulations with large noise (both on simulators and on a real device)
we find that the expectation value of a conserved charge~$Q$ of the integrable model decays exponentially and approaches constant values:
\begin{equation}\label{eq:Qn-asymptotics}
\langle Q\rangle_d  \sim  c_1 e^{-\gamma d} +  c_2  \quad \textrm{($d$  large)} \,,
\end{equation}
where $\gamma>0$ and $c_1, c_2\in \mathbb{R}$.
Moreover, the asymptotic value $c_2$ is independent of the choice of the initial state $\rho_0$.
Such a pattern is clear especially in our noisy simulations on classical emulators shown in FIG.~\ref{fig:depo-damp}.

We can explain this behavior as follows.
In our quantum simulation of the integrable circuit, we prepare an $N$-qubit density matrix $\rho_0 \sim |\psi_0\rangle \langle \psi_0|$, 
and repeatedly  apply a quantum channel $\Phi$ (i.e.\ a completely positive trace-preserving (CPTP) map) corresponding to a single Trotter step,
so that the expectation value after $d$ Trotter steps
\begin{equation}
\langle Q\rangle_{d} = {\rm tr}[ \Phi^{d}(\rho_0) \, Q]  \,,
\end{equation}
where $\Phi^d$ is the $d$-th power of $\Phi$.
From the theory of quantum channels (see e.g.\ \cite{Watrous,wolf2012quantum}),
it is known that all the eigenvalues $\{\lambda_n\}$ of $\Phi$ must have absolute values less than or equal to $1$,
and that $\Phi$ has $\lambda_0 = 1$ as an eigenvalue (i.e.\ $\Phi$ has a fixed point).
Denoting the fixed-point density matrix by $\rho_*$,
we have the asymptotic behavior~(\ref{eq:Qn-asymptotics}) with 
\begin{equation}
c_2 = \ {\rm tr}[\rho_* Q] \,.
\end{equation}
If we assume that there is no degeneracy in the eigenvalue $\lambda_0$ (as one expects in generic noisy situations),\footnote{In integrable models
one expects multiple steady states due to the existence of conserved charges \cite{Albert_2014}. In our setting, however, we expect the integrability to be broken due to noise and that no charges remain conserved.}
then $\rho_*$ is determined by the quantum channel only and is independent of the choice of the initial state.

\bigskip
The asymptotic density matrix $\rho_*$ depends heavily on the properties of the noise.
For the time evolution~$\Phi$ with depolarizing errors discussed in~\ref{sec:depo}, it is easy to see that $\Phi$ has the completely-mixed state $\rho_*=(1/2^N) I^{\otimes N}$ as a fixed point.\footnote{The completely-mixed state is invariant under $\Phi_{\rm depo}$ and any unitary transformations.}
This gives $c_2=0$ for traceless conserved charges $Q$, which is consistent with the behaviors of $\langle Q\rangle$ in FIG.~\ref{fig:depo-damp}. 
The situation is different for the time evolution with the amplitude-and-phase damping errors discussed in~\ref{sec:damp},\footnote{The completely-mixed state is also a fixed point if the error is purely from the phase damping ($\lambda_p\ne 0, \lambda_a=0$ in \eqref{eq:damp_Kraus}).}
where the asymptotic values of the charges do not necessarily vanish.
Although statistical uncertainties are larger, our simulation results from the real quantum device
are consistent with the hypothesis that the asymptotic values of $\langle Q_1^+\rangle$ and $\langle Q_1^\text{dif}\rangle$ both vanish. 
In all our simulations, the fact that the asymptotic values do not depend on the choice of the initial state indicates that the quantum channel has a unique fixed point.

\begin{figure}[t]
\begin{center}
\includegraphics[scale=0.4]{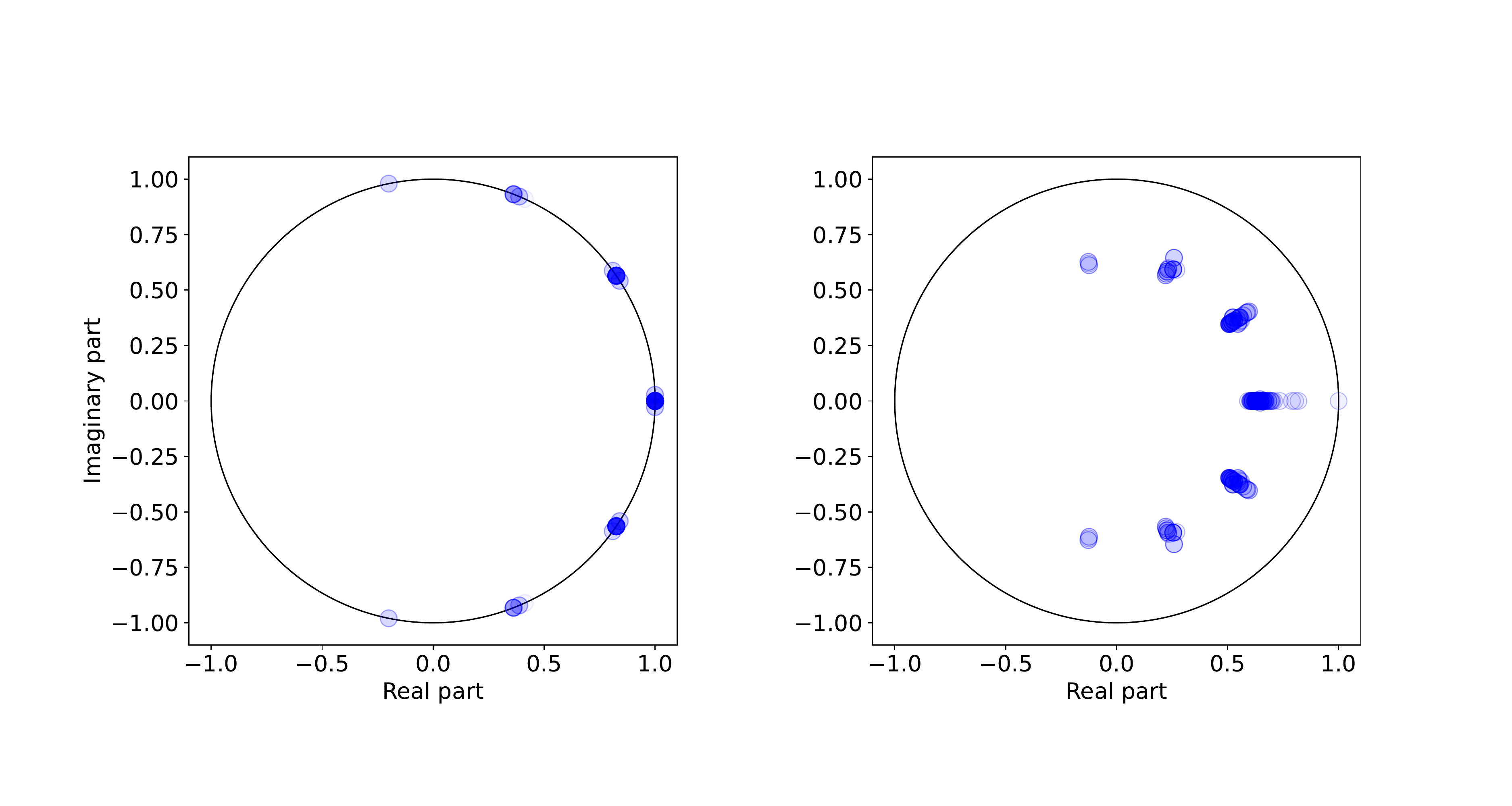}
\caption{Eigenvalue distributions of the quantum channel $\Phi$ representing the single-step time evolution on 4 sites, shown here for
(a) the noiseless (unitary) time evolution, and 
(b) the noisy time evolution.
As for the noisy time evolution (b), we exemplified the case of the depolarizing error with the error rate $p=0.018$.
Each figure has 256 ($=2^{2\times4}$) eigenvalues
and some of them are degenerate. 
A pale blue dot represents an eigenvalue; thus, the deeper colored dots indicate a larger degeneracy of the eigenvalues.
In particular, the eigenvalue $\lambda = 1$ appears with no degeneracy in the noisy case.
}
\label{fig:spectrum_circle}
\end{center}
\end{figure}

Whether the fixed point of $\Phi$ is genuinely unique is an important question. 
To clarify this point, we calculated the spectrum of the quantum channel $\Phi$ numerically (in a classical computer) on the system with 4 sites.
The details of this computation are explained in Appendix \ref{app:spectrum-details}.
FIG.\ref{fig:spectrum_circle} illustrates the eigenvalue distribution of the quantum channel $\Phi$ of noiseless and noisy (depolarizing error with the error rate $p = 0.018$) time evolution.\footnote{From the general properties of the CPTP map, we expect that
the spectrum $\textrm{Spec}(\Phi)$ should (1) contain $\lambda=1$ (2) be preserved by the complex conjugation
(3) $|\lambda| \le 1$ for all $\lambda \in \textrm{Spec}(\Phi)$.} 
The eigenvalues of the unitary channel are distributed on the unit circle. 
In contrast, all the eigenvalues of the noisy channel are inside the unit circle, except a single eigenvalue with the value 1.
As we increase the Trotter step in the noisy time evolution, the eigenvalues except for 1 move more deeply toward the center.

Let us make more quantitative analysis on the value of the decay rate $\gamma$ in \eqref{eq:Qn-asymptotics}.
One expects that the decay under the time evolution
is governed by the eigenvalue $\lambda=\lambda_1$ with 
the largest among $|\lambda|<1$, and that the late-time decay rate $\gamma$ (inverse of the relaxation time) of the charge should equal $-\ln|\lambda_1|$.%
\footnote{Importance of $|\lambda_1|$ has been discussed e.g.\ in \cite{Poulin_2010, Kessler_2012, H_ning_2012} for continuous-time master equations.}
While this relation should hold in idealized situations (at very late times and for $\lambda_1$ isolated from other eigenvalues), we find that it does not strictly hold for our simulation results.
Indeed, the values of $\gamma$ obtained by fitting the data in FIG.~\ref{fig:depo-damp}(a) summarized as 
\begin{equation}
\begin{array}{c|ccccc}
\text{charge}     &Q_1^+ & Q_1^\text{dif} & Q_2^+ & Q_2^\text{dif} & Q_3^+  \\
\hline
\gamma     &0.26&0.38&0.29&0.39&0.30
\end{array}    
\end{equation}
are comparable with but somewhat larger than  $-\ln|\lambda_1|\sim 0.20$, and 
depend on the type of the charge.
This difference appears because we obtained $\gamma$ by fitting the data for all steps $d$,
and also because as shown in FIG.~\ref{fig:spectrum_circle} we have many eigenvalues $\lambda$ whose absolute values are smaller but close to $|\lambda_1|$ --- this means that we will have a sum of exponentials
with comparable sizes.

\subsection{Towards benchmarking protocols from early-time decays}
\label{sec:benchmarking}

As we discussed in Section~\ref{sec:classical-sim-small-noise}, we expect that future real-device simulations with small noise rates will exhibit the linear behavior (\ref{eq:Q_linear_early}) for  the initial change of the charge $Q$.
The parameter $\beta$ in (\ref{eq:Q_linear_early}) quantifies how ``noisy'' the quantum device is.
It is therefore natural to propose to use $\beta$
as a benchmark for quantum devices in the future.

While there are already many benchmarking proposals in the literature,
there are some motivations for proposing a new benchmark. 
First, since we are discussing spin chains this benchmark could be 
more suitable for evaluating the performance of quantum many-body simulations.
Second, integrable models exhibit a characteristic complexity growth different from chaotic or free models \cite{Parker:2018yvk}.
In this sense, our benchmark is conceptually 
distinct from the randomized benchmarks related to chaotic time evolution \cite{2005JOptB...7S.347E,2007Sci...317.1893E,PhysRevA.77.012307,PhysRevLett.106.180504,2019PhRvA.100c2328C}.
Third, the higher conserved charges $\{ Q_n^\pm \}$ can be regarded as a large number of benchmark indicators, which provides refined data about the precision of real devices.

Note that exactly-known values of some particular quantities (e.g.\ energy as an eigenvalue of the Hamiltonian)
have often been used in calibrating quantum devices, and there are also papers on 
integrable-model benchmarks in the NISQ era \cite{2020arXiv200301862D,Cervia:2020fkk,2021PhRvA.104b2412R}.
However, the essence of integrability is not only about the solvability of the spectrum of the Hamiltonian,
but also the existence of as many conserved charges as the degrees of freedom. 
It is therefore natural to 
take advantage of all such conserved charges in the benchmarking.

Let us make some preliminary comments on the benchmarking protocols.
First, we should keep in mind that the parameter $\beta$ 
depends on many parameters/data:
the choice of the initial state $|\psi_0\rangle$, 
the size $\delta$ of the Trotter step,
the choice of the charge $Q$ (i.e.\ $Q=Q_n^{\pm}$ or $Q_n^{\rm diff}$ for some positive integer $n$),
the number of qubits (width $w$),
and the number of Trotterized time evolution (depth $d$).
While $\delta$ is a parameter independent from others,
for the purpose of simulating the real-time dynamics of the spin chain
it is natural to fix the product $d \cdot \delta$ to a constant value.

We can then propose the following procedure:\footnote{For precise specification of the benchmarks we need to specify the compilation rules:
we can either allow or forbid compilation across different copies of $\mathcal{U}(\delta)$.}
\begin{itemize}
   \item Fix the choice of the $w$-qubit initial state $|\psi_0\rangle $, as well as the threshold value $\beta_{\star}$ (say 1\%).
    \item Run the depth-$d$ quantum circuit that approximately prepares the state
    $\mathcal{U}(\delta)^d|\psi_0\rangle$.
    \item Compute $\langle Q\rangle$ by (many copies of) the state obtained above,
    and estimate the values of $\beta$ in \eqref{eq:Q_linear_early}.
    \item We declare pass/fail depending on whether or not $\beta$ is smaller/larger than $\beta_{\star}$. 
   We obtain a multi-dimensional plot by collecting the results for different values of $w, d, n$ and for different initial conditions.  
\end{itemize}

Compared with the ``volumetric benchmarks'' \cite{Blume_Kohout_2020} with width $w$ and depth $d$, here we have the third direction given by the order $n$ of the conserved charge, and the fourth direction given by the choice of the initial state.
The charges $Q_n$ are more non-local as we increase $n$ and could thus in principle be better at capturing more non-localized errors than the usual Hamiltonian. It remains to be seen if this new direction is of any use for benchmarking purposes.

\section{Conclusion and discussion}\label{sec:discussion}

In this work we performed the digital quantum simulation of the integrably-Trotterized spin-1/2 Heisenberg XXX spin chain with periodic boundary conditions on a real quantum computer and on classical simulators.
We designed and implemented the protocols to simulate the discrete time evolution and measurements. We computed the expectation values of the conserved charges together with their statistical uncertainties.
While integrability guarantees that the conserved charges remain constant under the time evolution, we found that quantum noise forces the charges to decay. 
We analyzed noise models and performed quantum state tomography on classical simulators, and confirmed that the decay of the conserved charges is a typical behavior of noisy systems.
Further, we demonstrated that the late-time behavior of the time evolution can be understood in terms of the spectral properties of the quantum channels.
We also proposed that the decay rate $\beta$ for the early-time behavior (\ref{eq:Q_linear_early})
in weakly noisy simulations can be used to benchmark quantum devices and algorithms in the future.
This benchmark setup is directly relevant to the simulation of quantum field theories.

We made modest theoretical progress regarding the integrable Trotterization of the XXX spin chain.
By rewriting the boost relations that recursively generate the conserved charges, we provided an efficient method to derive explicit expressions of the conserved charges using a classical computer program.
Based on these results, we obtained a simple expression (\ref{eq:q3+}) for the density of $Q_3^+$, as well as explicit expressions for higher charges up to $Q_6^+$. It would be interesting to find a closed-form expression for all $Q_n^+$, by generalizing the known results for the original XXX spin chain \cite{Grabowski:1994ae}.

While we focused on the spin-$1/2$ Heisenberg XXX spin chain in this paper, it is straightforward to
generalize our method to other integrable models.
In Section~\ref{sec:classical-sim-small-noise} we found that the effects of noise are different for different charges.
This may be due to the wider interaction range for the density of $Q_n^+$ with higher $n$.
Similarly, it is natural to ask if the effects of noise are different 
for different integrable models.
Turning this around, we may hope that a systematic analysis of noises in quantum devices could lead to new classifications of
quantum integrable models.

There are several points where our simulation methods may be improved.
While we made sufficiently many measurements (shots) for the simulation on {\tt ibm\_kawasaki}, 
we should be able to make our protocols and job management more efficient, thereby scaling up the qubit size $N$ and Trotter step $d$.
This will allow us to compute charges higher than $Q_1^+$ and $Q_1^{\rm dif}$ that we computed on a real device.
As a related issue, it is desirable to have a more quantitative understanding of the number of necessary measurement bases (Pauli words) needed to compute higher charges.
To alleviate the systematic errors noted in Section~\ref{sec:real-devices} and discussed in Appendix~\ref{app:simulation-details}, it is desirable to have available a hardware-level technique such as the Floquet calibration used in~\cite{2020arXiv201007965A}.
For simulation results in section~\ref{sec:results} we omitted error mitigation (see \cite{Endo_review} for a recent review), which we expect should alleviate the effects of noise.
As a preliminary step, we implemented two error mitigation methods on {\tt ibm\_kawasaki} for $Q_1^+$ on 4 sites and obtained improved results as summarized in Appendix \ref{app:MEM}.

In the continuum limit $\delta \rightarrow 0$, the time evolution with an initial state such as  the N\'eel state can be regarded as quantum quench.
See, for example, \cite{Calabrese:2006rx}.
When devices with lower noise rates are available, our protocols together with error mitigation will allow us to study
quantum quench via digital quantum simulation.

\begin{acknowledgments}
This work grew out of a quantum simulation study group (organized by Y.~Hidaka, TO
and MY in 2021), and we would like to thank all the participants there for stimulating discussions. 
The work of KM is supported in part by JSPS Grant-in-Aid for Scientific Research
Grant Number JP20K03935.
The work of TO and YY was supported in part by the JSPS Grant-in-Aid for Scientific Research No.\ 21H05190.
The work of MY is supported in part by the JSPS Grant-in-Aid for Scientific Research No.\ 17KK0087, 19K03820, 19H00689, 20H05860.
The work of JWP is supported in part by the JSPS Grant-in-Aid for Research Fellow No.\ 22J14732.
The work of RS is supported by NSFC grant no. 12050410255, and he thanks Yunfeng Jiang for stimulating discussions.
The work of YY is supported in part by JSPS Grant-in-Aid for Scientific Research Grant Number 21K03382.
This work is partly supported by UTokyo Quantum Initiative.
We thank IonQ for providing access to their quantum computers, and are grateful to Matt Keesan for technical help.

\

\noindent
{\bf Public availability of code and data:}
We make publicly available some programs and real-device data in our public GitHub repository \cite{repo}.
A Mathematica notebook creates conserved charge densities $q^{[n, \pm]}_{2j,2j+1, \dots, 2j+2n+2}$ up to $n=6$ using the method in Appendix \ref{sec:recursion}. 
We also provide simplified Python code that implements our protocol (state initialization, time evolution, and measurements) on a classical simulator.


\end{acknowledgments}

\appendix

\section{Recursion relations for conserved charges}
\label{sec:recursion}

In this appendix, we study in more detail the recursion relations for the conserved charges $Q^\pm_n$ discussed in \ref{subsec:charges}.

In terms of the densities $q^{[n \, \pm]}$ in \eqref{eq:Qpm} and \eqref{eq:Qpm-qpm}, the recursion relations \eqref{eq:boost} for $Q^{\pm}_n$ reads
\begin{equation} \label{eq:recursion_plus-explicit}
\left[
\sum_{\ell= - \infty}^{\infty} \ell \, \mathbb{R}'_{2\ell-3,2\ell-2|2 \ell-1,2\ell}(0)
, 
\sum_{j=-\infty}^{\infty} q^{[n\, \pm]}_{2j,2j+1,\ldots,2j+2n} 
\right] = 
\sum_{j=-\infty}^{\infty} q^{[n+1, \pm ]}_{2j,2j+1,\ldots,2j+2n+2}.
\end{equation}
To use the relation \eqref{eq:recursion_plus-explicit}, we assume that the densities take the form
\begin{equation}
q_{1\ldots 2n+1}^{[n\pm]} = c_{\mu_1\ldots\mu_{2n+1}}^{[n \pm]} \sigma_1^{\mu_1}\ldots \sigma_{2n+1}^{\mu_{2n+1}} \,,
\end{equation}
where the indices $\mu_1,\ldots,\mu_{2n+1}$ are summed over the values $0,1,2,3$, and $\sigma_a^0$ denotes the identity $I$ on site $a$.
Since 
the density is a polynomial in $\delta$, we define the coefficients in 
front of $\delta^m$ as
\begin{eqnarray}
 c_{\mu_1\ldots\mu_{2n+1}}^{[n \pm]}
  =     \sum_{m=0}^{2n} \delta^m c_{\mu_1\ldots\mu_{2n+1}}^{[n \pm] (m)}; ~~~~~
 q_{1\ldots 2n+1}^{[n\pm]} 
  =     \sum_{m=0}^{2n} \delta^m  q_{1\ldots 2n+1}^{[n\pm] (m)} .
 \end{eqnarray}
 These coefficients can be read off by
 \begin{equation}
c_{\mu_1\ldots \mu_{2n+1}}^{[n \pm](m)} = \frac{1}{2^{2n+1}} {\rm tr} \left(
\sigma_1^{\mu_1}\ldots \sigma_{2n+1}^{\mu_{2n+1}} q^{[n\pm](m)}_{1,2,\ldots,2n+1}
\right) \,,
 \end{equation}
where the trace is taken over $2n+1$ qubits.

There is one subtlety we have not touched upon:
given $Q_n^\pm$, there are ambiguities in the choices of the quantities $q^{[n \pm]}_{1,2,\ldots,2n+1}$ in \eqref{eq:Qpm} and \eqref{eq:Qpm-qpm},
coming from the simultaneous shift $\{ \sigma_j \} \to \{ \sigma_{j+2} \}$.
We can fix this ambiguity by requiring that $q^{[n\pm]}_{12\ldots2n+1}$ has no term that has $I$ (the identity) both at $2n$ and $2n+1$.
This means $c_{\mu_1\ldots\mu_{2n-1}00}^{[n \pm]} =0$\footnote{The same convention of the densities has been used in \cite{2018PhRvL.121c0606V}.}.

In terms of $q^{[n \pm](m)}$ expanded in $\delta$, the boost relation is written as
\begin{eqnarray}
&&\quad 
\sum_{k \in\mathbb Z}  q^{[n+1,+](m)}_{2k,\ldots,2k+2n+2} 
\nonumber\\
&&=
\frac{i}{2}\sum_{k\in\mathbb Z}
\Bigg(
\sum_{j=k+1}^{k+n+1}
2j\Big[
\bm{\sigma}_{2j-2} \cdot \bm\sigma_{2j-1}
  \ , \  
 q^{[n,+](m)}_{2k,\ldots,2k+2n} \Big]
 +\sum_{j=k}^{k+n}
 (2j+1) \Big[
 \bm{\sigma}_{2j-1} \cdot \bm\sigma_{2j}
  \ , \  
 q^{[n,+](m)}_{2k,\ldots,2k+2n} \Big]
 \nonumber\\
&& +\sum_{j=k+1}^{k+n+1}
 j \Big[
 \bm\sigma_{2j-3}\cdot ( \bm\sigma_{2j-2}\times  \bm\sigma_{2j-1})
     \ , \  
 q^{[n,+](m-1)}_{2k,\ldots,2k+2n}  
 \Big]
 -\sum_{j=k}^{k+n+1}
 j \Big[
 \bm\sigma_{2j-2}\cdot ( \bm\sigma_{2j-1}\times  \bm\sigma_{2j})
     \ , \  
 q^{[n,+](m-1)}_{2k,\ldots,2k+2n}  
 \Big]
  \nonumber\\
&& +\sum_{j=k+1}^{k+n+1}
j \Big[
\bm\sigma_{2j-3}\cdot\bm\sigma_{2j-1}
     \ , \  
 q^{[n,+](m-2)}_{2k,\ldots,2k+2n}  
 \Big]
+
\sum_{j=k}^{k+n+1}
j \Big[
\bm\sigma_{2j-2}\cdot\bm\sigma_{2j}
     \ , \  
 q^{[n,+](m-2)}_{2k,\ldots,2k+2n}  
 \Big]
 \Bigg) \,.\label{boost-q-hat}
\end{eqnarray}
We can translate this to the relations between the coefficients $c^{[n \pm](m)}$.
\footnotesize
  \begin{eqnarray}
  && c^{[n+1, +](m)}_{\nu_0 \ldots \nu_{2n} kl}
        \nonumber \\
  &&
   =     \sum_{j=2}^{n} 2 j \delta^0_{\nu_0} \delta^0_{\nu_1} f_{\nu_{2j-2} \nu_{2j-1}}^{\rho_1 \rho_2} c^{[n, +](m)}_{\nu_2 \ldots \nu_{2j-3} \rho_1 \rho_2 \nu_{2j} \ldots \nu_{2n}kl} 
        + 2(n+1) \delta^0_{\nu_0} \delta^0_{\nu_1} f_{\nu_{2n} k}^{\rho_1 \rho_2} c^{[n, +](m)}_{\nu_2 \ldots \nu_{2n-1} \rho_1 \rho_2 l}  
       + 3 \delta_{\nu_0}^0 f_{\nu_1 \nu_2}^{0 \rho} c^{[n, +](m)}_{\rho \nu_3 \ldots \nu_{2n} kl}
         \nonumber \\
  &&  + \sum_{j=2}^{n}  (2j+1) \delta^0_{\nu_0} \delta^0_{\nu_1} f_{\nu_{2j-1} \nu_{2j}}^{\rho_1 \rho_2} c^{[n, +](m)}_{\nu_2 \ldots \nu_{2j-2} \rho_1 \rho_2 \nu_{2j+1} \ldots \nu_{2n}kl} 
         + (2n+3) \delta^0_{\nu_0} \delta^0_{\nu_1} f_{kl}^{\rho_1 \rho_2} c^{[n, +](m)}_{\nu_2 \ldots \nu_{2n} \rho_1 \rho_2} 
           \nonumber \\
  && + 2(n+2) \delta^0_{\nu_0} \delta^0_{\nu_1}\delta^0_{\nu_2} \delta^0_{\nu_3} \epsilon^{\rho_1 \rho_2 l} c^{[n, +](m)}_{\nu_4 \ldots \nu_{2n} k \rho_1 \rho_2 0} 
           \nonumber \\
  &&
        + 2 \delta_{\nu_0}^0 g_{\nu_1 \nu_2 \nu_3}^{0 \rho_1 \rho_2} c_{\rho_1 \rho_2 \nu_4 \ldots \nu_{2n} kl}^{[n,+](m-1)}
        + \sum_{j=3}^n j \delta_{\nu_0}^0 \delta_{\nu_1}^0g_{\nu_{2j-3} \nu_{2j-2} \nu_{2j-1}}^{\rho_1 \rho_2 \rho_3} c_{\nu_2 \ldots \nu_{2j-4} \rho_1 \rho_2 \rho_3 \nu_{2j} \ldots \nu_{2n} kl}^{[n,+](m-1)}
          \nonumber \\
  &&
        + (n+1) \delta_{\nu_0}^0 \delta_{\nu_1}^0 g_{\nu_{2n-1} \nu_{2n} k}^{\rho_1 \rho_2 \rho_3} c_{\nu_2 \ldots \nu_{2n-2} \rho_1 \rho_2 \rho_3 l}^{[n,+](m-1)}
        +(n+2) \delta^0_{\nu_0} \delta^0_{\nu_1} \delta^0_{\nu_2} \delta^0_{\nu_3 } f_{kl}^{\gamma_3 | \gamma_1 \gamma_2} c^{[n,+](m-1)}_{\nu_4\ldots \nu_{2n} \gamma_1 \gamma_2 \gamma_3 0} 
        \nonumber \\
  &&
       - (n+1)  g_{\nu_{2n} kl}^{\rho 00} c^{[n,+](m-1)}_{\nu_0 \ldots \nu_{2n-1} \rho}
       - g_{\nu_0 \nu_1 \nu_2}^{00\rho}  c^{[n,+](m-1)}_{\rho \nu_3 \ldots \nu_{2n} kl}
       - \sum_{j=2}^n j \delta^0_{\nu_0} \delta^0_{\nu_1} g_{\nu_{2j-2} \nu_{2j-1} \nu_{2j}}^{\rho_1 \rho_2 \rho_3}  c^{[n,+](m-1)}_{\nu_2 \ldots \nu_{2j-3} \rho_1 \rho_2 \rho_3 \nu_{2j+1} \ldots \nu_{2n} kl}
         \nonumber \\
  &&
       - (n+1) \delta^0_{\nu_0} \delta^0_{\nu_1} g_{\nu_{2n} kl}^{\rho_1 \rho_2 \rho_3}  c^{[n,+](m-1)}_{\nu_2 \ldots \nu_{2n-1} \rho_1 \rho_2 \rho_3 }
       +(n+2) \delta^0_{\nu_0} \delta^0_{\nu_1} \delta^0_{\nu_2} \delta^0_{\nu_3 } (c^{[n,+](m-1)}_{\nu_4\ldots \nu_{2n} k \rho_1 \rho_1 l} - c^{[n,+](m-1)}_{\nu_4\ldots \nu_{2n} k \rho_1 l \rho_1})
       \nonumber \\
  &&
        + 2 \delta_{\nu_0}^0 f^{0 \rho}_{\nu_1 \nu_3} c^{[n,+](m-2)}_{\nu_2 \rho \nu_4\ldots \nu_{2n}kl} 
        + \sum_{j=3}^n j \delta_{\nu_0}^0 \delta_{\nu_1}^0 f^{\rho_1 \rho_2}_{\nu_{2j-3} \nu_{2j-1}} c^{[n,+](m-2)}_{\nu_2 \ldots \nu_{2j-4} \rho_1 \nu_{2j-2} \rho_2 \nu_{2j} \ldots \nu_{2n}kl} 
        + (n+1) \delta_{\nu_0}^0 \delta_{\nu_1}^0 f^{\rho_1 \rho_2}_{\nu_{2n-1} k} c^{[n,+](m-2)}_{\nu_2 \ldots \nu_{2n-2} \rho_1 \nu_{2n} \rho_2 l} 
        \nonumber \\
  && + (n+2) \delta_{\nu_0}^0 \delta_{\nu_1}^0 \delta_{\nu_2}^0 \delta_{\nu_3}^0 \epsilon^{\rho_1 \rho_2 k} c^{[n,+](m-2)}_{\nu_4 \ldots \nu_{2n} \rho_1 l \rho_2 0} 
        \nonumber \\
  &&
        + \delta_{\nu_1}^0 f^{0 \rho}_{\nu_{0} \nu_2} c^{[n,+](m-2)}_{\rho \nu_3 \ldots \nu_{2n} kl}
       + \sum_{j=2}^n j \delta_{\nu_0}^0 \delta_{\nu_1}^0 f^{\rho_1 \rho_2}_{\nu_{2j-2} \nu_{2j}} c^{[n,+](m-2)}_{\nu_2 \ldots \nu_{2j-3} \rho_1 \nu_{2j-1} \rho_2 \nu_{2j+1} \ldots \nu_{2n}kl} 
        + (n+1) \delta_{\nu_0}^0 \delta_{\nu_1}^0 f^{\rho_1 \rho_2}_{\nu_{2n} l} c^{[n,+](m-2)}_{\nu_2 \ldots \nu_{2n-1} \rho_1 k \rho_2} 
        \nonumber \\
  && + (n+2) \delta_{\nu_0}^0 \delta_{\nu_1}^0 \delta_{\nu_2}^0 \delta_{\nu_3}^0 \epsilon^{\rho_1 \rho_2 l} c^{[n,+](m-2)}_{\nu_4 \ldots \nu_{2n} k \rho_1 0 \rho_2} ~,
  \label{eq:recursion1}
  \end{eqnarray}%
  \begin{eqnarray}
  && c^{[n+1, +](m)}_{\nu_0 \ldots \nu_{2n} 0l}
        \nonumber \\
  &&
   =    \sum_{j=2}^{n} 2 j \delta^0_{\nu_0} \delta^0_{\nu_1} f_{\nu_{2j-2} \nu_{2j-1}}^{\rho_1 \rho_2} c^{[n, +](m)}_{\nu_2 \ldots \nu_{2j-3} \rho_1 \rho_2 \nu_{2j} \ldots \nu_{2n}0l} 
        + 2(n+1) \delta^0_{\nu_0} \delta^0_{\nu_1} f_{\nu_{2n} 0}^{\rho_1 \rho_2} c^{[n, +](m)}_{\nu_2 \ldots \nu_{2n-1} \rho_1 \rho_2 l}   
        + 3 \delta_{\nu_0}^0 f_{\nu_1 \nu_2}^{0 \rho} c^{[n, +](m)}_{\rho \nu_3 \ldots \nu_{2n} 0l}
           \nonumber \\
  &&  + \sum_{j=2}^{n}  (2j+1) \delta^0_{\nu_0} \delta^0_{\nu_1} f_{\nu_{2j-1} \nu_{2j}}^{\rho_1 \rho_2} c^{[n, +](m)}_{\nu_2 \ldots \nu_{2j-2} \rho_1 \rho_2 \nu_{2j+1} \ldots \nu_{2n}0l} 
         + (2n+3) \delta^0_{\nu_0} \delta^0_{\nu_1} f_{0l}^{\rho_1 \rho_2} c^{[n, +](m)}_{\nu_2 \ldots \nu_{2n} \rho_1 \rho_2} 
         \nonumber \\
  && + 2(n+2) \delta^0_{\nu_0} \delta^0_{\nu_1}\delta^0_{\nu_2} \delta^0_{\nu_3} \epsilon^{\rho_1 \rho_2 l} c^{[n, +](m)}_{\nu_4 \ldots \nu_{2n} 0 \rho_1 \rho_2 0} 
           \nonumber \\
  &&
        + 2 \delta_{\nu_0}^0 g_{\nu_1 \nu_2 \nu_3}^{0 \rho_1 \rho_2} c_{\rho_1 \rho_2 \nu_4 \ldots \nu_{2n} 0l}^{[n,+](m-1)}
        + \sum_{j=3}^n j \delta_{\nu_0}^0 \delta_{\nu_1}^0 g_{\nu_{2j-3} \nu_{2j-2} \nu_{2j-1}}^{\rho_1 \rho_2 \rho_3} c_{\nu_2 \ldots \nu_{2j-4} \rho_1 \rho_2 \rho_3 \nu_{2j} \ldots \nu_{2n} 0l}^{[n,+](m-1)}
          \nonumber \\
  &&
        + (n+1) \delta_{\nu_0}^0 \delta_{\nu_1}^0 g_{\nu_{2n-1} \nu_{2n} 0}^{\rho_1 \rho_2 \rho_3} c_{\nu_2 \ldots \nu_{2n-2} \rho_1 \rho_2 \rho_3 l}^{[n,+](m-1)}
        +(n+2) \delta^0_{\nu_0} \delta^0_{\nu_1} \delta^0_{\nu_2} \delta^0_{\nu_3 } f_{0l}^{\gamma_3 | \gamma_1 \gamma_2} c^{[n,+](m-1)}_{\nu_4\ldots \nu_{2n} \gamma_1 \gamma_2 \gamma_3 0} 
        \nonumber \\
  &&
       - (n+1)  g_{\nu_{2n} 0l}^{\rho_1 00} c^{[n,+](m-1)}_{\nu_0 \ldots \nu_{2n-1} \rho}
       - g_{\nu_0 \nu_1 \nu_2}^{00\rho}  c^{[n,+](m-1)}_{\rho \nu_3 \ldots \nu_{2n} 0l}
       - \sum_{j=2}^n j \delta^0_{\nu_0} \delta^0_{\nu_1} g_{\nu_{2j-2} \nu_{2j-1} \nu_{2j}}^{\rho_1 \rho_2 \rho_3}  c^{[n,+](m-1)}_{\nu_2 \ldots \nu_{2j-3} \rho_1 \rho_2 \rho_3 \nu_{2j+1} \ldots \nu_{2n} 0l}
        \nonumber \\
  &&
       - (n+1) \delta^0_{\nu_0} \delta^0_{\nu_1} g_{\nu_{2n} 0l}^{\rho_1 \rho_2 \rho_3}  c^{[n,+](m-1)}_{\nu_2 \ldots \nu_{2n-1} \rho_1 \rho_2 \rho_3 }
       +(n+2) \delta^0_{\nu_0} \delta^0_{\nu_1} \delta^0_{\nu_2} \delta^0_{\nu_3 } (c^{[n,+](m-1)}_{\nu_4\ldots \nu_{2n} 0 \rho_1 \rho_1 l} - c^{[n,+](m-1)}_{\nu_4\ldots \nu_{2n}  0 \rho_1 l \rho_1})
        \nonumber \\
  &&
        + 2 \delta_{\nu_0}^0 f^{0 \rho}_{\nu_1 \nu_3} c^{[n,+](m-2)}_{\nu_2 \rho \nu_4\ldots \nu_{2n}0l} 
        + \sum_{j=3}^n j \delta_{\nu_0}^0 \delta_{\nu_1}^0 f^{\rho_1 \rho_2}_{\nu_{2j-3} \nu_{2j-1}} c^{[n,+](m-2)}_{\nu_2 \ldots \nu_{2j-4} \rho_1 \nu_{2j-2} \rho_2 \nu_{2j} \ldots \nu_{2n}0l} 
        + (n+1) \delta_{\nu_0}^0 \delta_{\nu_1}^0 f^{\rho_1 \rho_2}_{\nu_{2n-1} 0} c^{[n,+](m-2)}_{\nu_2 \ldots \nu_{2n-2} \rho_1 \nu_{2n} \rho_2 l} 
        \nonumber \\
  &&
        + (n+1) f^{\rho 0}_{\nu_{2n} l} c^{[n,+](m-2)}_{\nu_0 \ldots \nu_{2n-1} \rho}
        \nonumber \\
  &&
        + \delta_{\nu_1}^0 f^{0 \rho}_{\nu_{0} \nu_2} c^{[n,+](m-2)}_{\rho \nu_3 \ldots \nu_{2n} 0l}
       + \sum_{j=2}^n j \delta_{\nu_0}^0 \delta_{\nu_1}^0 f^{\rho_1 \rho_2}_{\nu_{2j-2} \nu_{2j}} c^{[n,+](m-2)}_{\nu_2 \ldots \nu_{2j-3} \rho_1 \nu_{2j-1} \rho_2 \nu_{2j+1} \ldots \nu_{2n}0l} 
        + (n+1) \delta_{\nu_0}^0 \delta_{\nu_1}^0 f^{\rho_1 \rho_2}_{\nu_{2n} l} c^{[n,+](m-2)}_{\nu_2 \ldots \nu_{2n-1} \rho_1 0 \rho_2} 
        \nonumber \\
  && + (n+2) \delta_{\nu_0}^0 \delta_{\nu_1}^0 \delta_{\nu_2}^0 \delta_{\nu_3}^0 \epsilon^{\rho_1 \rho_2 l} c^{[n,+](m-2)}_{\nu_4 \ldots \nu_{2n} 0 \rho_1 0 \rho_2} ~,
  \end{eqnarray}
  
  \begin{eqnarray}
  &&    c^{[n+1, +](m)}_{\nu_0 \ldots \nu_{2n} k0}
          \nonumber \\
  &&
   =    2(n+1) f_{\nu_{2n} k}^{\rho_1 0} c_{\nu_0 \ldots \nu_{2n-1} \rho_1}
        + \sum_{j=2}^{n} 2 j \delta^0_{\nu_0} \delta^0_{\nu_1} f_{\nu_{2j-2} \nu_{2j-1}}^{\rho_1 \rho_2} c^{[n, +](m)}_{\nu_2 \ldots \nu_{2j-3} \rho_1 \rho_2 \nu_{2j} \ldots \nu_{2n}k0} 
        + 2(n+1) \delta^0_{\nu_0} \delta^0_{\nu_1} f_{\nu_{2n} k}^{\rho_1 \rho_2} c^{[n, +](m)}_{\nu_2 \ldots \nu_{2n-1} \rho_1 \rho_2 0} 
          \nonumber \\
  &&     
       + 3 \delta_{\nu_0}^0 f_{\nu_1 \nu_2}^{0 \rho} c^{[n, +](m)}_{\rho \nu_3 \ldots \nu_{2n} k0}
       + \sum_{j=2}^{n}  (2j+1) \delta^0_{\nu_0} \delta^0_{\nu_1} f_{\nu_{2j-1} \nu_{2j}}^{\rho_1 \rho_2} c^{[n, +](m)}_{\nu_2 \ldots \nu_{2j-2} \rho_1 \rho_2 \nu_{2j+1} \ldots \nu_{2n} k0} 
       + (2n+3) \delta^0_{\nu_0} \delta^0_{\nu_1} f_{k0}^{\rho_1 \rho_2} c^{[n, +](m)}_{\nu_2 \ldots \nu_{2n} \rho_1 \rho_2}
       \nonumber \\
  &&
        + (n+1) g_{\nu_{2n-1} \nu_{2n} k}^{\rho_1 \rho_2 0} c_{\nu_0 \ldots \nu_{2n-2}\rho_1 \rho_2}^{[n,+](m-1)}
        + 2 \delta_{\nu_0}^0 g_{\nu_1 \nu_2 \nu_3}^{0 \rho_1 \rho_2} c_{\rho_1 \rho_2 \nu_4 \ldots \nu_{2n} k0}^{[n,+](m-1)}
        + \sum_{j=3}^n j \delta_{\nu_0}^0 \delta_{\nu_1}^0 g_{\nu_{2j-3} \nu_{2j-2} \nu_{2j-1}}^{\rho_1 \rho_2 \rho_3} c_{\nu_2 \ldots \nu_{2j-4} \rho_1 \rho_2 \rho_3 \nu_{2j} \ldots \nu_{2n} k0}^{[n,+](m-1)}
          \nonumber \\
  &&
        + (n+1) \delta_{\nu_0}^0 \delta_{\nu_1}^0 g_{\nu_{2n-1} \nu_{2n} k}^{\rho_1 \rho_2 \rho_3} c_{\nu_2 \ldots \nu_{2n-2} \rho_1 \rho_2 \rho_3 0}^{[n,+](m-1)}
        +(n+2) \delta^0_{\nu_0} \delta^0_{\nu_1} \delta^0_{\nu_2} \delta^0_{\nu_3 } f_{k0}^{\gamma_3 | \gamma_1 \gamma_2} c^{[n,+](m-1)}_{\nu_4\ldots \nu_{2n} \gamma_1 \gamma_2 \gamma_3 0} 
        \nonumber \\
  &&
       - (n+1)  g_{\nu_{2n} k0}^{\rho_1 00} c^{[n,+](m-1)}_{\nu_0 \ldots \nu_{2n-1} \rho}
       - g_{\nu_0 \nu_1 \nu_2}^{00\rho}  c^{[n,+](m-1)}_{\rho \nu_3 \ldots \nu_{2n} k0}
       - \sum_{j=2}^n j \delta^0_{\nu_0} \delta^0_{\nu_1} g_{\nu_{2j-2} \nu_{2j-1} \nu_{2j}}^{\rho_1 \rho_2 \rho_3}  c^{[n,+](m-1)}_{\nu_2 \ldots \nu_{2j-3} \rho_1 \rho_2 \rho_3 \nu_{2j+1} \ldots \nu_{2n} k0}
         \nonumber \\
  &&
       - (n+1) \delta^0_{\nu_0} \delta^0_{\nu_1} g_{\nu_{2n} k0}^{\rho_1 \rho_2 \rho_3}  c^{[n,+](m-1)}_{\nu_2 \ldots \nu_{2n-1} \rho_1 \rho_2 \rho_3 }
       +(n+2) \delta^0_{\nu_0} \delta^0_{\nu_1} \delta^0_{\nu_2} \delta^0_{\nu_3 } (c^{[n,+](m-1)}_{\nu_4\ldots k \rho_1 \rho_1 0} - c^{[n,+](m-1)}_{\nu_4\ldots k \rho_1 0 \rho_1})
        \nonumber\\
  &&
        + (n+1) f^{\rho 0}_{\nu_{2n-1} k} c^{[n,+](m-2)}_{\nu_0 \ldots \nu_{2n-2} \rho \nu_{2n}} 
        \nonumber \\
  &&
        + 2 \delta_{\nu_0}^0 f^{0 \rho}_{\nu_1 \nu_3} c^{[n,+](m-2)}_{\nu_2 \rho \nu_4\ldots \nu_{2n-1}k0} 
        + \sum_{j=3}^n j \delta_{\nu_0}^0 \delta_{\nu_1}^0 f^{\rho_1 \rho_2}_{\nu_{2j-3} \nu_{2j-1}} c^{[n,+](m-2)}_{\nu_2 \ldots \nu_{2j-4} \rho_1 \nu_{2j-2} \rho_2 \nu_{2j} \ldots \nu_{2n}k0} 
        + (n+1) \delta_{\nu_0}^0 \delta_{\nu_1}^0 f^{\rho_1 \rho_2}_{\nu_{2n-1} k} c^{[n,+](m-2)}_{\nu_2 \ldots \nu_{2n-2} \rho_1 \nu_{2n} \rho_2 0} 
        \nonumber \\
  &&
        + (n+2) \delta_{\nu_0}^0 \delta_{\nu_1}^0 \delta_{\nu_2}^0 \delta_{\nu_3}^0 \epsilon^{\rho_1 \rho_2 k} c^{n,+}(m-2)_{\nu_4 \ldots \nu_{2n} \rho_1 0 \rho_2 0} 
        \nonumber \\
  &&
        + \delta_{\nu_1}^0 f^{0 \rho}_{\nu_{0} \nu_2} c^{[n,+](m-2)}_{\rho \nu_3 \ldots \nu_{2n} k0}
       + \sum_{j=2}^n j \delta_{\nu_0}^0 \delta_{\nu_1}^0 f^{\rho_1 \rho_2}_{\nu_{2j-2} \nu_{2j}} c^{[n,+](m-2)}_{\nu_2 \ldots \nu_{2j-3} \rho_1 \nu_{2j-1} \rho_2 \nu_{2j+1} \ldots \nu_{2n}k0} 
        + (n+1) \delta_{\nu_0}^0 \delta_{\nu_1}^0 f^{\rho_1 \rho_2}_{\nu_{2n} 0} c^{[n,+](m-2)}_{\nu_2 \ldots \nu_{2n-1} \rho_1 k \rho_2} ~,
        \label{eq:recursion3}
  \end{eqnarray}
  \normalsize
where we have defined
\footnotesize
  \begin{eqnarray}
  f_{\nu_1 \nu_2}^{\gamma | \mu_1 \mu_2}
  &=&    \frac{i}{8} \epsilon^{\gamma \rho_1 \rho_2} {\rm Tr}_{1,2} (\sigma_1^{\nu_1} \sigma_2^{\nu_2} [ \sigma_1^{\rho_1} \sigma_2^{\rho_2}, \sigma_1^{\mu_1} \sigma_2^{\mu_2} ] ),
  \nonumber \\
  f^{\mu_1\mu_2}_{\nu_1\nu_2} 
  &=& \frac{i}{8}
{\rm tr}_{12}\left(
\sigma_{1}^{\nu_{1}}\sigma_2^{\nu_{2}}
\left[
\bm{\sigma}_{1} \cdot \bm\sigma_{2}
  \ , \  
 \sigma_{1}^{\mu_{1}}\sigma_{2}^{\mu_{2}}
\right]\right) \,,
 \nonumber \\
  g^{\mu_1\mu_2\mu_3}_{\nu_1\nu_2\nu_3} 
  &=& \frac{i}{16}
{\rm tr}_{123}\left(
\sigma_{1}^{\nu_{1}}\sigma_2^{\nu_{2}}\sigma_3^{\nu_{3}}
\left[
\bm{\sigma}_{1} \cdot (\bm\sigma_{2} \times \bm\sigma_{3})
  \ , \  
 \sigma_{1}^{\mu_{1}}\sigma_{2}^{\mu_{2}}\sigma_{3}^{\mu_{3}}
\right]\right) \,.
  \end{eqnarray}
  \normalsize
We have $f^{\mu_1\mu_2}_{\nu_1\nu_2}  =(-1)^{\delta_{\nu_1}^{0}+\delta_{\nu_2}^{0}}  \epsilon^{\mu_1\mu_2\nu_1\nu_2}$,
where $\epsilon$ is a completely-antisymmetric tensor with $\epsilon^{0123}=1$.

These equations determine the coefficients recursively,
from which one can get the higher order densities; {\it e.g.} $q^{[3,+]}$ is given by
\footnotesize
  \begin{eqnarray}
    q^{[3, +]}_{1,2,3,4,5,6,7}
    &=& -4 {\boldsymbol\sigma}_6\cdot{\boldsymbol\sigma}_7 + 2{\boldsymbol\sigma}_5\cdot{\boldsymbol\sigma}_7 -4 {\boldsymbol\sigma}_5\cdot{\boldsymbol\sigma}_6 + 2{\boldsymbol\sigma}_4\cdot{\boldsymbol\sigma}_6 + 2{\boldsymbol\sigma}_4\cdot ({\boldsymbol\sigma}_5 \times {\boldsymbol\sigma}_6 \times {\boldsymbol\sigma}_7) + 2{\boldsymbol\sigma}_3\cdot ({\boldsymbol\sigma}_4 \times {\boldsymbol\sigma}_5 \times {\boldsymbol\sigma}_6) 
    \nonumber \\
    & &  + \delta \Big(
    10 {\boldsymbol\sigma}_5\cdot ({\boldsymbol\sigma}_6 \times {\boldsymbol\sigma}_7) - 2 {\boldsymbol\sigma}_4\cdot ({\boldsymbol\sigma}_6 \times {\boldsymbol\sigma}_7) - 4 {\boldsymbol\sigma}_4\cdot ({\boldsymbol\sigma}_5 \times {\boldsymbol\sigma}_7) + 8 {\boldsymbol\sigma}_4\cdot ({\boldsymbol\sigma}_5 \times {\boldsymbol\sigma}_6) - 4 {\boldsymbol\sigma}_3\cdot ({\boldsymbol\sigma}_5 \times {\boldsymbol\sigma}_6) 
    \nonumber \\
    & & 
    - 2 {\boldsymbol\sigma}_3\cdot ({\boldsymbol\sigma}_4 \times {\boldsymbol\sigma}_6) - 4 {\boldsymbol\sigma}_3\cdot ({\boldsymbol\sigma}_4 \times {\boldsymbol\sigma}_5 \times {\boldsymbol\sigma}_6 \times {\boldsymbol\sigma}_7)- 2 {\boldsymbol\sigma}_2\cdot ({\boldsymbol\sigma}_3 \times {\boldsymbol\sigma}_4 \times {\boldsymbol\sigma}_5 \times {\boldsymbol\sigma}_6)
    \Big)
    \nonumber \\
    & & + \delta^2 \Big(
    2 {\boldsymbol\sigma}_6\cdot{\boldsymbol\sigma}_7 - 10 {\boldsymbol\sigma}_5\cdot{\boldsymbol\sigma}_7 + 2 {\boldsymbol\sigma}_5\cdot{\boldsymbol\sigma}_6 + 2 {\boldsymbol\sigma}_4\cdot{\boldsymbol\sigma}_7 + 2 {\boldsymbol\sigma}_4\cdot{\boldsymbol\sigma}_6 + 2 {\boldsymbol\sigma}_3\cdot{\boldsymbol\sigma}_6 
    \nonumber \\
    & & -6 {\boldsymbol\sigma}_4\cdot ({\boldsymbol\sigma}_5 \times {\boldsymbol\sigma}_6 \times {\boldsymbol\sigma}_7) + 6 {\boldsymbol\sigma}_3\cdot ({\boldsymbol\sigma}_5 \times {\boldsymbol\sigma}_6 \times {\boldsymbol\sigma}_7) + 2 {\boldsymbol\sigma}_3\cdot ({\boldsymbol\sigma}_4 \times {\boldsymbol\sigma}_6 \times {\boldsymbol\sigma}_7) + 6 {\boldsymbol\sigma}_3\cdot ({\boldsymbol\sigma}_4 \times {\boldsymbol\sigma}_5 \times {\boldsymbol\sigma}_7) 
    \nonumber \\
    & &
    - 6 {\boldsymbol\sigma}_3\cdot ({\boldsymbol\sigma}_4 \times {\boldsymbol\sigma}_5 \times {\boldsymbol\sigma}_6) + 2 {\boldsymbol\sigma}_2\cdot ({\boldsymbol\sigma}_3 \times {\boldsymbol\sigma}_5 \times {\boldsymbol\sigma}_6) 
    \nonumber \\
    & &
    + 2 {\boldsymbol\sigma}_2\cdot ({\boldsymbol\sigma}_3 \times {\boldsymbol\sigma}_4 \times {\boldsymbol\sigma}_5 \times {\boldsymbol\sigma}_6 \times {\boldsymbol\sigma}_7) + 2 {\boldsymbol\sigma}_1\cdot ({\boldsymbol\sigma}_2 \times {\boldsymbol\sigma}_3 \times {\boldsymbol\sigma}_4 \times {\boldsymbol\sigma}_5 \times {\boldsymbol\sigma}_6) 
    \Big)
    \nonumber \\
    & & + \delta^3 \Big(
    6 {\boldsymbol\sigma}_5\cdot ({\boldsymbol\sigma}_6 \times {\boldsymbol\sigma}_7) - 2 {\boldsymbol\sigma}_4\cdot ({\boldsymbol\sigma}_6 \times {\boldsymbol\sigma}_7) + 4 {\boldsymbol\sigma}_4\cdot ({\boldsymbol\sigma}_5 \times {\boldsymbol\sigma}_7) - 2 {\boldsymbol\sigma}_3\cdot ({\boldsymbol\sigma}_6 \times {\boldsymbol\sigma}_7) - 8 {\boldsymbol\sigma}_3\cdot ({\boldsymbol\sigma}_5 \times {\boldsymbol\sigma}_7) 
    \nonumber \\
    & &
    - 2 {\boldsymbol\sigma}_3\cdot ({\boldsymbol\sigma}_4 \times {\boldsymbol\sigma}_6) + 4 {\boldsymbol\sigma}_3\cdot ({\boldsymbol\sigma}_5 \times {\boldsymbol\sigma}_6) - 2 {\boldsymbol\sigma}_3\cdot ({\boldsymbol\sigma}_4 \times {\boldsymbol\sigma}_7) + 4 {\boldsymbol\sigma}_3\cdot ({\boldsymbol\sigma}_4 \times {\boldsymbol\sigma}_5 \times {\boldsymbol\sigma}_6 \times {\boldsymbol\sigma}_7) 
    \nonumber \\
    & &
    - 2 {\boldsymbol\sigma}_2\cdot ({\boldsymbol\sigma}_3 \times {\boldsymbol\sigma}_5 \times {\boldsymbol\sigma}_6 \times {\boldsymbol\sigma}_7) - 2 {\boldsymbol\sigma}_2\cdot ({\boldsymbol\sigma}_3 \times {\boldsymbol\sigma}_4 \times {\boldsymbol\sigma}_5 \times {\boldsymbol\sigma}_7) - 2 {\boldsymbol\sigma}_1\cdot ({\boldsymbol\sigma}_3 \times {\boldsymbol\sigma}_4 \times {\boldsymbol\sigma}_5 \times {\boldsymbol\sigma}_6)  
    \nonumber \\
    & & 
    - 2 {\boldsymbol\sigma}_1\cdot ({\boldsymbol\sigma}_2 \times {\boldsymbol\sigma}_3 \times {\boldsymbol\sigma}_5 \times {\boldsymbol\sigma}_6) - 2  {\boldsymbol\sigma}_1\cdot ({\boldsymbol\sigma}_2 \times{\boldsymbol\sigma}_3 \times {\boldsymbol\sigma}_4 \times {\boldsymbol\sigma}_5 \times {\boldsymbol\sigma}_6 \times {\boldsymbol\sigma}_7) \Big)
    \nonumber \\
    & & + \delta^4 \Big( 
    - 2 {\boldsymbol\sigma}_6\cdot{\boldsymbol\sigma}_7 - 8 {\boldsymbol\sigma}_5\cdot{\boldsymbol\sigma}_7 - 2 {\boldsymbol\sigma}_5\cdot{\boldsymbol\sigma}_6 + 2 {\boldsymbol\sigma}_4\cdot{\boldsymbol\sigma}_7 + 2 {\boldsymbol\sigma}_3\cdot{\boldsymbol\sigma}_6 + 2 {\boldsymbol\sigma}_3\cdot{\boldsymbol\sigma}_7 - 2 {\boldsymbol\sigma}_3\cdot ({\boldsymbol\sigma}_5 \times {\boldsymbol\sigma}_6 \times {\boldsymbol\sigma}_7) 
    \nonumber \\
    & &
    + 2 {\boldsymbol\sigma}_3\cdot ({\boldsymbol\sigma}_4 \times {\boldsymbol\sigma}_6 \times {\boldsymbol\sigma}_7) - 2 {\boldsymbol\sigma}_3\cdot ({\boldsymbol\sigma}_4 \times {\boldsymbol\sigma}_5 \times {\boldsymbol\sigma}_7) + 2  {\boldsymbol\sigma}_2\cdot ({\boldsymbol\sigma}_3 \times {\boldsymbol\sigma}_5 \times {\boldsymbol\sigma}_7) + 2  {\boldsymbol\sigma}_1\cdot ({\boldsymbol\sigma}_3 \times {\boldsymbol\sigma}_5 \times {\boldsymbol\sigma}_6) 
    \nonumber \\
    & &
    + 2 {\boldsymbol\sigma}_1\cdot ({\boldsymbol\sigma}_3 \times {\boldsymbol\sigma}_4 \times {\boldsymbol\sigma}_5 \times {\boldsymbol\sigma}_6 \times {\boldsymbol\sigma}_7) + 2 {\boldsymbol\sigma}_1\cdot ({\boldsymbol\sigma}_2 \times {\boldsymbol\sigma}_3 \times {\boldsymbol\sigma}_5 \times {\boldsymbol\sigma}_6 \times {\boldsymbol\sigma}_7) + 2 {\boldsymbol\sigma}_1\cdot ({\boldsymbol\sigma}_2 \times {\boldsymbol\sigma}_3 \times {\boldsymbol\sigma}_4 \times {\boldsymbol\sigma}_5 \times {\boldsymbol\sigma}_7)
    \Big)
    \nonumber \\
    & & + \delta^5 \Big(
    4 {\boldsymbol\sigma}_5\cdot ({\boldsymbol\sigma}_6 \times {\boldsymbol\sigma}_7) - 2 {\boldsymbol\sigma}_3 \cdot ({\boldsymbol\sigma}_6 \times {\boldsymbol\sigma}_7) - 2 {\boldsymbol\sigma}_3\cdot ({\boldsymbol\sigma}_4 \times {\boldsymbol\sigma}_7) -2 {\boldsymbol\sigma}_1 \cdot ({\boldsymbol\sigma}_3 \times {\boldsymbol\sigma}_5 \times {\boldsymbol\sigma}_6 \times {\boldsymbol\sigma}_7)
    \nonumber \\
    & &
    -2 {\boldsymbol\sigma}_1 \cdot ({\boldsymbol\sigma}_3 \times {\boldsymbol\sigma}_4 \times {\boldsymbol\sigma}_5 \times {\boldsymbol\sigma}_7) -2 {\boldsymbol\sigma}_1 \cdot ({\boldsymbol\sigma}_2 \times {\boldsymbol\sigma}_3 \times {\boldsymbol\sigma}_5 \times {\boldsymbol\sigma}_7)
    \Big)
    \nonumber \\
    & & + \delta^6 \Big( 
    - 4 {\boldsymbol\sigma}_5\cdot{\boldsymbol\sigma}_7 + 2 {\boldsymbol\sigma}_3\cdot{\boldsymbol\sigma}_7 + 2 {\boldsymbol\sigma}_1\cdot ( {\boldsymbol\sigma}_3 \times {\boldsymbol\sigma}_5 \times {\boldsymbol\sigma}_7)
    \Big).
    \label{eq:q3+}
  \end{eqnarray}
\normalsize

We generated the conserved charge densities up to $n=6$ using this method. The explicit results for $n \ge 4$ can be found in~\cite{repo}.
They suggest that the number of terms in $q^{[n,+]}$ increases exponentially with~$n$.

\section{Formulas for statistical uncertainties}
\label{app:stat-unc-est}

Following definitions, it is straightforward to check that the expectation value  $\mathbb{E}[\langle Q\rangle_{\rm est}]$ (with respect to the probability distribution defined by many copies of the quantum state $\rho$) of the estimator~$\langle Q\rangle_{\rm est}$ given in~(\ref{eq:estimator-Q}) indeed equals $\langle Q\rangle \equiv {\rm tr}(\rho Q)$.

In practice, we also need to estimate the statistical uncertainty in the estimator $\langle Q\rangle_{\rm est}$.
The estimator is computed from the outcomes of many measurements in various measurement bases.
If we repeat this set of measurements many times, the value of the estimator itself fluctuates according to a probability distribution whose variance is, by definition
\begin{align}
&\quad \mathbb{E}[(\langle Q\rangle_{\rm est} - \mathbb{E}[\langle Q\rangle_{\rm est} ]) ^2] 
=
\sum_{P,P'}\sum_{W\supset P}\sum_{W'\supset P'}\sum_{i=1}^{n_W}\sum_{i'=1}^{n_{W'}}\frac{c_{ Q,P}c_{ Q,P'}}{n_P n_{P'}}\mathbb{E}\left[\left(
 \Pi_{P,W,i} - \langle P\rangle\right)\left(
  \Pi_{P',W',i'} 
  - \langle P'\rangle\right) \right]\,,\label{eq:Q-est-var2}
\end{align}
where
\begin{equation}
 \Pi_{P,W,i} 
  := \prod_{\mytop{j=1}{w_{j}(P) \neq I}}^{N}(-1)^{b_{j}^{(W, i)}}.
\end{equation}
The summand in~(\ref{eq:Q-est-var2}) vanishes unless $(W,i)=(W',i')$.
By expanding the product we obtain
\begin{align}
\quad \mathbb{E}[(\langle Q\rangle_{\rm est} - \mathbb{E}[\langle Q\rangle_{\rm est} ]) ^2] 
&=
\sum_{P,P'}\sum_{W\supset P,P'}\sum_{i=1}^{n_W}\frac{c_{ Q,P}c_{ Q,P'}}{n_P n_{P'}}
\left(
\mathbb{E}\left[
\Pi_{P,W,i} 
\Pi_{P',W,i} 
\right]
-
\langle P \rangle\langle P'\rangle
\right)
\nonumber\\
&=
\sum_{P,P'} 
\frac{n_{P,P'}}{n_P n_{P'}} c_{ Q,P}c_{ Q,P'}
\left(
\langle P P'\rangle
-
\langle P \rangle\langle P'\rangle
\right) \,, \label{eq:Q-est-var}
\end{align}
where $n_{P,P'}:=\sum_{W\supset P,P'}n_W$.
If the quantum state $\rho$ is known explicitly,
the expression (\ref{eq:Q-est-var}) can be easily computed.
But we should also be able to estimate the variance from measurement results.
To find a good estimator, let us define
\begin{equation}
  S_{P,P'}: = \frac{1}{n_{P,P^\prime}}
  \sum_{W \supset P,  P^{\prime}}
  \sum_{i=1}^{n_{W}}\left(
   \Pi_{P,W,i} -\langle P\rangle_{\mathrm{est}}^{P^{\prime}}\right)
  \left(
   \Pi_{P',W,i} 
   -\left\langle P^{\prime}\right\rangle_{\mathrm{est}}^{P}\right),
\end{equation}
where
\begin{equation}
  \langle P\rangle_{\text {est }}^{P^{\prime}}:=\frac{1}{n_{P,P^\prime}} \sum_{W \supset P, P^{\prime}} 
  \Pi_{P,W,i} \,.
\end{equation}
Writing $\Pi$ for $ \Pi_{P,W,i} $ and $\Pi'$ for $ \Pi_{P',W,i} $ for brevity,
we have
\begin{eqnarray}
  \mathbb{E}\left[ S_{P,P'} \right]
  &=&\mathbb{E}\left[ 
    \frac{1}{n_{P ,P^{\prime}}} \sum_{W \supset P, P^{\prime}}\sum_{i=1}^{n_{W}}
  \left\{
    \left(\Pi-\langle P\rangle\right) 
    - \left(\langle P\rangle_{\mathrm{est}}^{P^{\prime}} - \langle P\rangle\right)\right\}
  \left\{
    \left(\Pi^{\prime} - \left\langle P^{\prime}\right\rangle \right)
    -\left(\left\langle P^{\prime}\right\rangle_{\mathrm{est}}^{P} - \left\langle P^{\prime}\right\rangle\right)
    \right\}
    \right]\\
  &=&
  \mathbb{E}\left[ \frac{1}{n_{P ,P^{\prime}}} \sum_{W \supset P, P^{\prime}}\sum_{i=1}^{n_{W}}
  \left(\Pi-\langle P\rangle\right) \left(\Pi^{\prime} - \left\langle P^{\prime}\right\rangle \right)
  \right]
  -\mathbb{E}\left[ \frac{1}{n_{P ,P^{\prime}}} \sum_{W \supset P, P^{\prime}}\sum_{i=1}^{n_{W}} 
  \left(\Pi-\langle P\rangle\right)
  \left(\left\langle P^{\prime}\right\rangle_{\mathrm{est}}^{P} - \left\langle P^{\prime}\right\rangle\right)
  \right]\nonumber\\
  &&
  -\mathbb{E}\left[ \frac{1}{n_{P ,P^{\prime}}} \sum_{W \supset P, P^{\prime}}\sum_{i=1}^{n_{W}}
  \left(\langle P\rangle_{\mathrm{est}}^{P^{\prime}} - \langle P\rangle\right)\left(\Pi^{\prime} - \left\langle P^{\prime}\right\rangle \right)
  \right]
  +\mathbb{E}\left[
  \left(\langle P\rangle_{\mathrm{est}}^{P^{\prime}} - \langle P\rangle\right)
  \left(\left\langle P^{\prime}\right\rangle_{\mathrm{est}}^{P} - \left\langle P^{\prime}\right\rangle\right)
  \right]\\
  &=& \mathbb{E}\left[ \frac{1}{n_{P ,P^{\prime}}} \sum_{W \supset P, P^{\prime}}\sum_{i=1}^{n_{W}}
  \left(\Pi-\langle P\rangle\right) \left(\Pi^{\prime} - \left\langle P^{\prime}\right\rangle \right)
  \right]-\mathbb{E}\left[
    \left(\langle P\rangle_{\mathrm{est}}^{P^{\prime}} - \langle P\rangle\right)
    \left(\left\langle P^{\prime}\right\rangle_{\mathrm{est}}^{P} - \left\langle P^{\prime}\right\rangle\right)
    \right] \,,
\end{eqnarray}
where we have used the relation
\begin{equation}
  \langle P\rangle_{\text {est }}^{P^{\prime}}-\langle P\rangle = \frac{1}{n_{P ,P^{\prime}}} \sum_{W \supset P, P^{\prime}}\sum_{i=1}^{n_{W}} 
  \left(\Pi_{P,W,i}-\langle P\rangle\right)  
\end{equation}
to obtain the last line.
Using the relation once again and evaluating the expectation values, we obtain
\begin{equation}
\mathbb{E}[S_{P,P'}]  = \left(1 - \frac{1}{n_{P,P^\prime}}\right)
  \left(\left\langle P P^{\prime}\right\rangle-\langle P\rangle\left\langle P^{\prime}\right\rangle \right).
\end{equation}
This shows that (\ref{eq:sQ2}) is an unbiased estimator of the variance~(\ref{eq:Q-est-var}):
\begin{equation}
    \mathbb{E}[s_Q^2] = 
\quad \mathbb{E}[(\langle Q\rangle_{\rm est} - \mathbb{E}[\langle Q\rangle_{\rm est} ]) ^2]  .
\end{equation}

\section{Details and more data in real-device simulations}
\label{app:simulation-details}

\subsection{Simulations on IBM devices}\label{app:IBM}

\begin{figure}[t]
\begin{center}
 \includegraphics[scale=0.25]{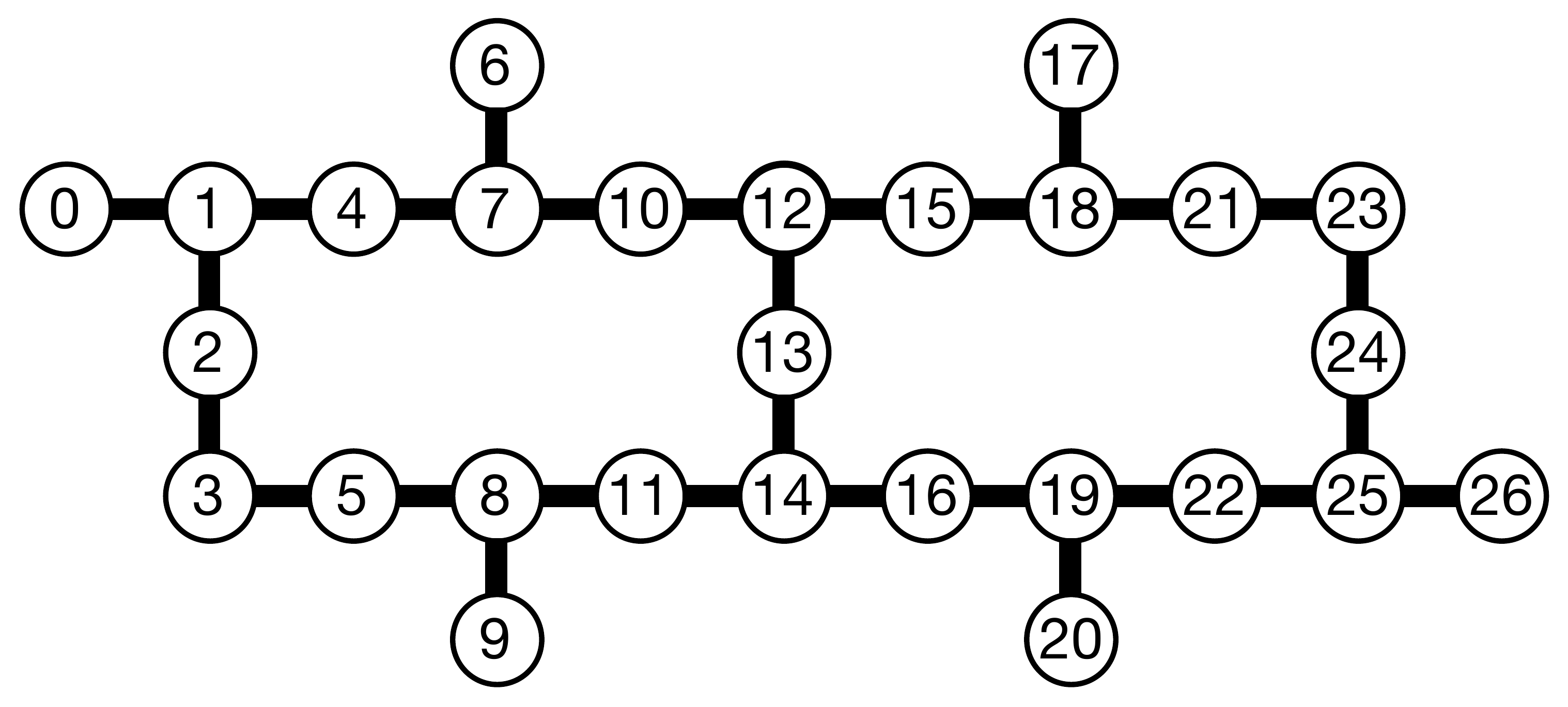}
\caption{Coupling of physical qubits on the {\tt ibm\_kawasaki} device.
Each circle represents a qubit.
}
\label{fig:IBM-kawasaki-coupling}
\end{center}
\end{figure}

FIG.~\ref{fig:IBM-kawasaki-coupling} shows the coupling of qubits on the {\tt ibm\_kawasaki} device (Falcon r5.11). 
Let us specify the mapping of logical qubits (sites~$j$ in Section~\ref{sec:staggered-XXX}) to physical qubits (nodes with numbers $q_j$ in FIG.~\ref{fig:IBM-kawasaki-coupling}) by a sequence $[q_0,q_1,\ldots,q_{N-1}]$, with the convention $q_0 = q_N$.
The results in FIG.~\ref{fig:IBM-kawasaki-results} were obtained with the qubit mappings
\begin{equation}
\begin{array}{r@{\quad:\quad}l}
\text{4 sites} & [12,13,14,16] \\
\text{6 sites}& [12,13,14,16,19,22] \\
\text{8 sites}& [12,13,14,16,19,22,25,24] \\ 
\text{10 sites}& [12,13,14,16,19,22,25,24,23,21] \\ 
\text{12 sites}& [12,13,14,16,19,22,25,24,23,21,18,15]
\end{array}
\end{equation}
both for $Q_1^+$ and  $Q_1^{\rm dif}$.
We chose the physical qubits whose connections have higher values of average fidelity for CNOT operations.
Note that for 12 sites the physical qubits form a loop on the device.
The CNOT gate fidelity, averaged over all the connections excluding the 11-14 connection, is $0.9933 = 1-6.70\times 10^{-3}$ 
according to the calibration data obtained from the IBM Quantum website.

\begin{figure}[t]
\begin{center}
\begin{tabular}{cc}
 \includegraphics[scale=0.7]{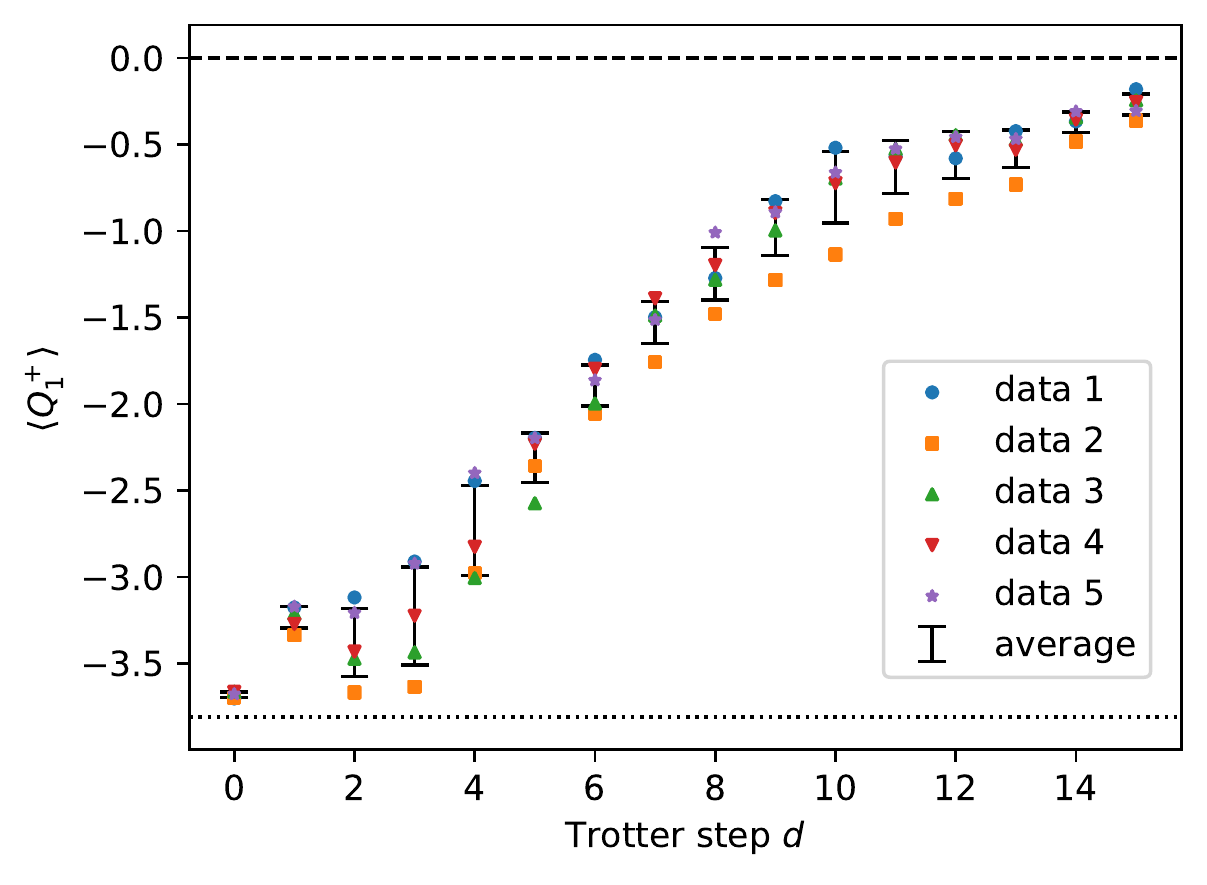}
 & \includegraphics[scale=0.7]{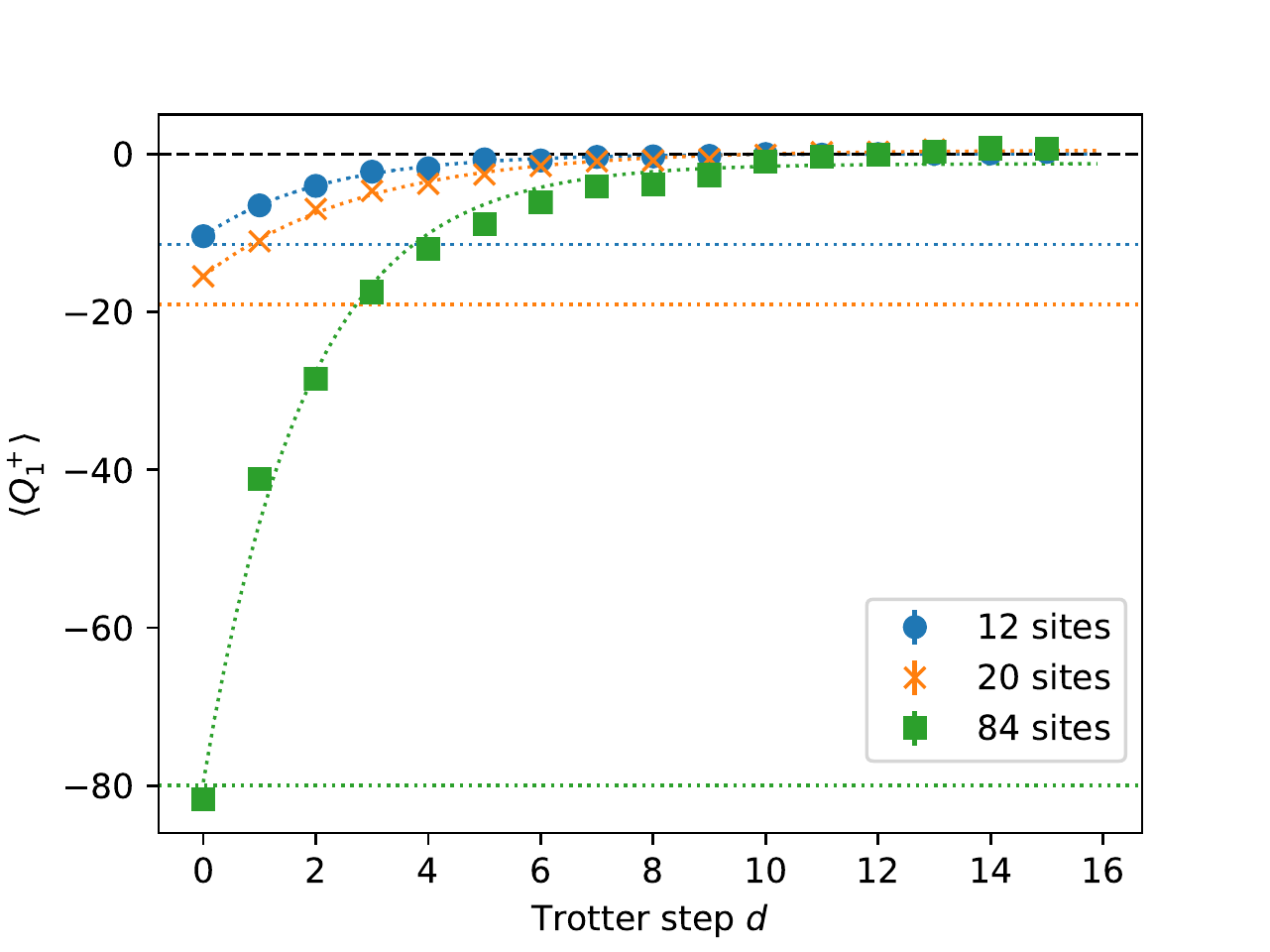}
 \\
(a) & (b)
\end{tabular}
\caption{(a) Results of 5 repeated simulations on {\tt ibm\_kawasaki} for $Q_1^+$ on 4 sites with  $\alpha=0.3$.
The error bars represent the standard deviation of the five data for each Trotter step $d$.
The dotted horizontal line indicates the value in the noiseless case.
Data 1 appears in FIG.~\ref{fig:IBM-kawasaki-results}(a) as the 4-site result.
(b) Simulation results on {\tt ibm\_washington} for $Q_1^+$ on 12, 20, and 84 sites with $\alpha=0.3$ and the initial state $|\text{N\'eel}\rangle =|01\ldots 01\rangle$.
Error bars are hidden by markers.
The expectation values in the noiseless case, represented by the dotted lines, are $-11.4$ for 12 sites, $-19.0$ for 20 sites, and $-80.0$ for 84 sites.
}
\label{fig:IBM_extra}
\end{center}
\end{figure}

The results in FIG.~\ref{fig:IBM-kawasaki-results} showed fluctuations as a function of Trotter step $d$, implying the existence of sizable systematic uncertainties.
To estimate them, we performed 5 repeated simulations (some days apart and some minutes apart) on {\tt ibm\_kawasaki} for $Q_1^+$ on 4 sites with $\alpha=0.3$ and the qubit mapping [12,13,14,16], with the results shown in FIG.~\ref{fig:IBM_extra}(a).
We note that the data for 4 sites have the largest fluctuations in  FIG.~\ref{fig:IBM-kawasaki-results}.

In addition to the {\tt ibm\_kawasaki} device, we ran simulations on {\tt ibm\_washington} (Eagle~r1) with 127 qubits.
The device has a configuration of physical qubits similar to but larger than the one in FIG.~\ref{fig:IBM-kawasaki-coupling}, and chose qubits that form circular loops.
We also avoided qubits and couplings that have high error rates.
In FIG.~\ref{fig:IBM_extra}(b), we show the 
results for {\tt ibm\_washington}.
They show behaviors qualitatively similar to those for {\tt ibm\_kawasaki}, but decay faster and do not fluctuate much from one step to the next.
The exponential decay behavior in {\tt ibm\_washington} does not significantly differ between 12 sites and 84 sites.

\subsection{Quantum simulations on IonQ devices}\label{app:IonQ}

\begin{figure}[t]
\begin{center}
 \includegraphics[scale=0.7]{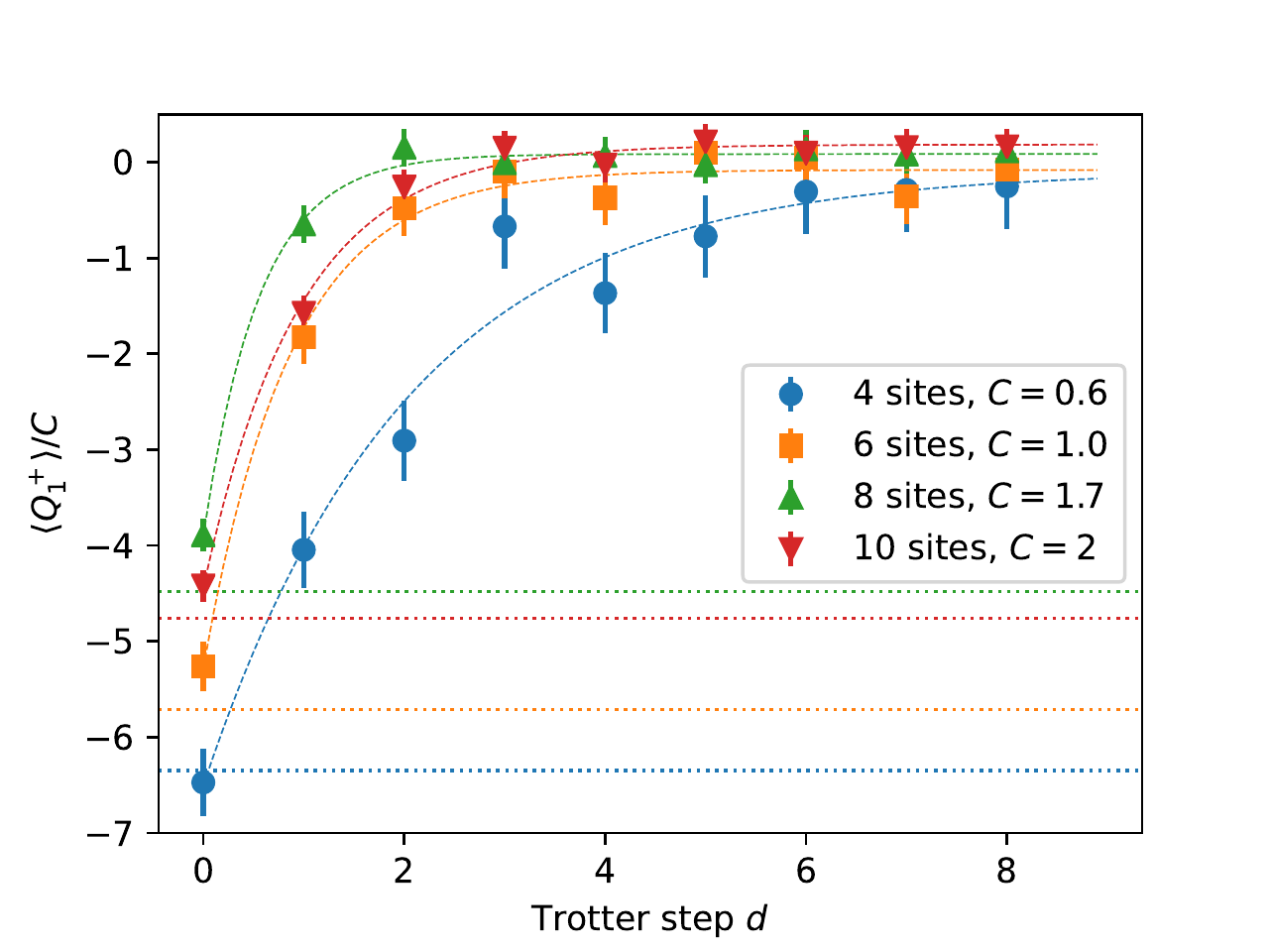}

\caption{
IonQ simulation results for charge $Q_1^+$ and the initial state $|\text{N\'eel}\rangle =|01\ldots 01\rangle$.
The total number of circuit executions (shots) for each $d$ was 2,250 for 4 sites, 2,650 for 6 sites, 2,600 for 8 sites, and 2,650 for 10 sites.
The error bars represent statistical uncertainties.
The expectation values in the noiseless case, represented by the dotted lines, are $-3.81$ for 4 sites, $-5.71$ for 6 sites, $-7.62$ for 8 sites, and $-9.52$ for 10 sites.
}
\label{fig:IonQ-6-10}
\end{center}
\end{figure}

Here we describe the quantum simulations performed on the trapped-ion quantum device {\tt Harmony} provided by IonQ. The device has 11 qubits
with all-to-all couplings. Since our model requires an even number of sites, the maximal number is 10. 
FIG.~\ref{fig:IonQ-6-10} shows
the simulation results for charge $Q_1^+$ and $Q_1^\text{dif}$ on 4, 6, 8, and 10 sites. As in the IBM case, the charges decay as a function of $d$.

\section{Quantum state tomography algorithm}\label{app:qst}

The density matrix of an $N$-qubit system is a $2^N \times 2^N$ positive semi-definite Hermitian matrix with unit trace, whose matrix element can be determined by $4^N-1$ real parameters. Such a matrix can be expanded in the Pauli basis as
\begin{equation}
\rho = \sum_{P \in \{I,X,Y,Z\}^{\otimes N}} a_{P} P  \qquad (a_P \in \mathbb{R}),
\label{def:rho by Pauli}
\end{equation}
with ${\rm tr} (PQ) = 2^N \delta_{PQ}$. We should set $a_P = 1$ for $P=I^{\otimes N}$ to solve the unit-trace condition.

Suppose that the measurement result of the observable $P$ is $m_P$\,, which may contain various statistical and systematic errors.
We consider the following weighted log-likelihood function,
\begin{equation}
{\cal L}_{\rm log} (\rho) = \sum_{P \in \{I,X,Y,Z\}^{\otimes N}} w_P \, \Bigl\{ m_P - {\rm tr} ( \rho P ) \Bigr\}^2  \qquad (w_P \ge 0).
\label{def:log-likelihood}
\end{equation}
This function ${\cal L}_{\rm log}$ has the global minimum at $\rho_* = \frac{1}{2^N} \, \sum_P m_P P$, which is called the linear inversion solution. This $\rho_*$ is a Hermitian matrix, but not always positive semi-definite for generic $\{ m_P \}$. We have to solve a convex optimization problem numerically to find a consistent density matrix.

If the weights are equal, $w_P=1 \ (\forall P)$, the log-likelihood function becomes\footnote{Use the identities $m_P = {\rm tr} (\rho_* P)$ and $\rho = \frac{1}{2^N} \, \sum_P {\rm tr} (\rho P) P$.}
\begin{equation}
{\cal L}_{\rm log} (\rho) = {\rm tr} \Bigl[ ( \rho_* - \rho )^2 \Bigr] .
\label{log-likelihood2}
\end{equation}
One can obtain the global minimum of this function in the space of density matrices by the PSD (positive semi-definite) rescaling \cite{Smolin_2012}.

\bigskip
The measurements for a mixed state can be described by a set of orthogonal projections
\begin{equation}
\Pi_i \, \Pi_ j = \delta_{ij} \ \Pi_i \,, \qquad
\sum_{i=0}^{2^N-1} \Pi_i = {\bf 1}_{2^N} \,,
\label{def:PVM}
\end{equation}
which is called a projection-valued measure (PVM). 
For example, the PVM in the computational basis is given by the binary representation of $i=0,1, \dots, 2^N-1$ as
\begin{equation}
\Pi_0 = \ket{0 \dots 00} \bra{0 \dots 00}, \quad
\Pi_1 = \ket{0 \dots 01} \bra{0 \dots 01}, \quad
\dots \quad
\Pi_{2^N-1} = \ket{1 \dots 11} \bra{1 \dots 11}.
\label{def:comp PVM}
\end{equation}
Each element $\Pi_i$ is the Kronecker product of the $i$-th eigenvector of $Z^{\otimes N}$ and its dual vector.
The measurements of $\{ \Pi_i \}$ gives $2^N$ integers,
\begin{equation}
{\rm tr} (\rho \, \Pi_i) \simeq \frac{s_i}{S} \,, \qquad 
\sum_{i=0}^{2^N-1} s_i = S, \qquad
(s_i \in \mathbb{Z}_{\ge 0}),
\end{equation}
where $S$ is the total number of shots.
The symbol $\simeq$ means that we estimate the LHS (a number) from the RHS (an integer-valued random variable).

Since a general density matrix \eqref{def:rho by Pauli} has $4^N-1$ real parameters, we need more PVM's for quantum state tomography.
We use the so-called Pauli measurement basis, which consists of $3^N$ PVM's,
\begin{equation}
{\rm PVM}^{(k)} = \{ \Pi_0^{(k)} , \Pi_1^{(k)} , \dots, \Pi_{2^N-1}^{(k)} \}, \qquad
(k=1,2, \dots, 3^N).
\label{def:Pauli PVM}
\end{equation}
Here $\Pi_i^{(k)}$ is the Kronecker product of the $i$-th eigenvector of ${\bf P}^{(k)}$ and its dual vector, where ${\bf P}^{(k)}$ is the $k$-th element of the tuples $\{ X, Y, Z \}^{\otimes N}$.\footnote{Such a ${\bf P}^{(k)}$ is called a Pauli word in Section \ref{sec:measure-Pauli}. For a fixed $k$, the $2^N$ elements in \eqref{def:Pauli PVM} are orthogonal with each other.} 
The measurements of $\{ \Pi_i^{(k)} \}$ gives $6^N$ integers,
\begin{equation}
{\rm tr} (\rho \, \Pi_i^{(k)} ) \simeq \frac{s_i^{(k)}}{S} \,, \qquad 
\sum_{i} s_i^{(k)} = S, \qquad
(k=1,2, \dots, 3^N),
\label{def:povm sik}
\end{equation}
where we assumed that the number of shots is the same for each measurement observable ${\bf P}^{(k)}$.
From the measurement results, we define the weighted log-likelihood function
\begin{equation}
{\cal L}_{\rm log}(\rho) = \sum_{k=1}^{3^N} \sum_{i=0}^{2^N-1} w_{k,i} \ \Bigl\{ p_{k,i} - {\rm tr} ( \rho \, \Pi_i^{(k)} ) \Bigr\}^2  , \qquad 
\left( p_{k,i} \equiv \frac{s_i^{(k)}}{S}  \right)
\label{log-likelihood PVM}
\end{equation}
and look for its (local) minimum in the space of density matrices. 
Note that this POVM basis $\{ \Pi_i^{(k)} \}$ does not satisfy the orthogonality condition ${\rm tr} (PQ) = 2^N \delta_{PQ}$\,.
Even if all the weights are trial $w_{k,i}=1$\,, it is not guaranteed that the PSD rescaling gives the global minimum of the log-likelihood function in the space of density matrices.

It is straightforward to construct the quantum circuits for the Pauli measurement basis \eqref{def:Pauli PVM}, because $(X, Y, Z)$ for each qubit can be measured as in FIG.~\ref{fig:circuits}(b).
However, the log-likelihood function \eqref{log-likelihood PVM} contains $6^N$ parameters whereas the density matrix contains $(4^N-1)$ real parameters as in \eqref{def:rho by Pauli}.
The Pauli measurement basis is over-complete for quantum state tomography, and the diagonal elements of $\rho$ are usually more often measured than off-diagonal elements.

\section{Details of spectral analysis} \label{app:spectrum-details}

For the spectral analysis in Section~\ref{subsec:early},
it is useful to invoke the Choi--Jamio{\l}kowski isomorphism (channel-state duality) and 
represent the quantum channel as an operator acting on the 
states $|\rho \rangle$ representing the density matrix $\rho$ \cite{CHOI1975285,JAMIOLKOWSKI1972275}.

The density matrix $\rho$ is a positive operator $\mathcal{H}\rightarrow\mathcal{H}$ acting on the Hilbert space $\mathcal{H}$,
which can be represented in an orthonormal basis $\{ |\alpha\rangle \} $ of the Hilbert space as
\begin{eqnarray}
 \rho &=& \sum_{\alpha, \beta} {\rho}_{\alpha \beta}|\alpha\rangle \otimes \langle \beta |. 
\end{eqnarray}
The vectorized state $|\rho\rangle$ of $\rho$ can then be defined as 
\begin{eqnarray}
  |\rho\rangle := \sum_{\alpha, \beta} {\rho}_{\alpha \beta}|\alpha\rangle \otimes|\beta\rangle.
\end{eqnarray}
In this representation, the time evolution of the 
density matrix $\rho$ by a quantum channel (CPTP map) $\Phi$ is represented by a positive operator $\widetilde{\Phi}: \mathcal{H} \otimes \mathcal{H} \rightarrow \mathcal{H} \otimes \mathcal{H}$ on $\mathcal{H} \otimes \mathcal{H}$ satisfying
\begin{eqnarray}
  \widetilde{\Phi}|\mathcal{\rho}\rangle=|\Phi(\mathcal{\rho})\rangle.
\end{eqnarray}

For our purposes, recall that the circuit for the noiseless single-step time evolution in Section 3.1
is given by 
\begin{align}\label{eq:noiseless_evolution}
  \Qcircuit @C=1em @R=1.4em {
  & \multigate{1}{
  R
  } & \qw \\ 
  & \ghost{R } & \qw
  } 
  \quad 
  &\raisebox{-4mm}{$=$} \quad 
  \Qcircuit @C=1em @R=.7em {
  & \ctrl{1} & \gate{H} & \ctrl{1} & \gate{R_Z (-\alpha)} & \ctrl{1} & \gate{H}   & \ctrl{1}
  & \qw  \\
  & \targ & \qw & \targ & \gate{R_Z (\alpha)} & \targ & \gate{R_Z (-\alpha)}  & \targ 
 & \qw 
  } 
\end{align}
We replace each unitary gate $U$ with an operator  $\widetilde{U} = U\otimes U^{*}$, e.g.
\begin{align}
  \Qcircuit @C=1em @R=1.4em {
  & \gate{H} & \qw \\ 
  &\qw& \qw
  } 
  \quad 
  &\raisebox{-3mm}{$\Longrightarrow \quad \widetilde{H\otimes\text{Id}} = \left( H \otimes \mathrm{Id} \right) \otimes \left( H \otimes \mathrm{Id} \right)^{*}$}
\end{align}

Let us include the effect of noises 
represented by a quantum channel $\Phi_{\rm err}$ acting on a single qubit.
In terms of the Kraus operators $\{ D_k\}$, this can be represented as, e.g.\
\begin{align}
  \Qcircuit @C=1em @R=1.4em {
  & \measure{\mbox{ $\Phi_{\rm err}$ }} & \qw \\ 
  &\qw& \qw
  } 
  \quad 
  &\raisebox{-3mm}{$\Longrightarrow\quad  \widetilde{\Phi_{\rm err}\otimes \mathrm{Id}} = \sum_k \left( D_k \otimes \mathrm{Id} \right) \otimes \left( D_k \otimes \mathrm{Id} \right)^{*}, $}
\end{align}
The noisy time evolution $\Phi$ is defined by 
inserting the quantum channel $\Phi_{\rm err}$ into the noiseless time evolution \eqref{eq:noiseless_evolution}:
\begin{align}
  \Qcircuit @C=1em @R=1.4em {
  & \multimeasure{1}{
  \Phi
  } & \qw \\ 
  & \ghost{\Phi} & \qw
  } 
  \quad 
  &\raisebox{-4mm}{$ = $} 
  \Qcircuit @C=1em @R=.7em {
  & \ctrl{1} & \measure{\mbox{ $\Phi_{\rm err}$ }}& \gate{H}& \measure{\mbox{ $\Phi_{\rm err}$ }} &  \ctrl{1} & \measure{\mbox{ $\Phi_{\rm err}$ }}&\qw &\cdots \\
  & \targ & \measure{\mbox{ $\Phi_{\rm err}$ }} & \qw & \qw& \targ & \measure{\mbox{ $\Phi_{\rm err}$ }}
 & \qw &\cdots 
  }  \raisebox{-4mm}{$ \quad \Longrightarrow \widetilde{\Phi} $}
\end{align}
The noisy time evolution at time $d$ is obtained by repeated action of the operator $\widetilde{\Phi}$
\begin{eqnarray}
  | \rho(d) \rangle = \widetilde{\Phi}^d | \rho_0 \rangle,
\end{eqnarray}
and the expectation value $\langle Q \rangle_d $ of the charge $Q$
is given by 
\begin{eqnarray}
  \langle Q \rangle_d = \langle Q | \rho(d)\rangle = \langle Q | \widetilde{\Phi}^d | \rho_0 \rangle.
 \end{eqnarray}

We studied the eigensystem of two types of $\widetilde{\Phi}$, one with the depolarizing error and the other with the damping errors.
Both errors are introduced in section \ref{subsec:large_noise}, and the corresponding Kraus operators are defined in \eqref{eq:dep_Kraus} and \eqref{eq:damp_Kraus}, respectively.
We confirmed that the analytic results calculated by the above algorithm agree with those from the quantum simulations using the noise model on the Qiskit for small values of $N$.

\section{Error  mitigation}\label{app:MEM}

In this appendix, we report our attempt to reduce the effects of quantum noise in simulation results using error mitigation techniques.
Specifically, we consider two typical error-mitigating algorithms: readout error mitigation~\cite{2019PhRvA.100e2315C,2019arXiv190708518M} and zero-noise extrapolation (ZNE) \cite{Li_2017_Mitigate,Temme_2017_Mitigate}. 
The simulation data in this appendix are distinct from those in the rest of the paper.
\par
The readout error mitigation technique mitigates the effect of quantum noise occurring at the measurement. 
This algorithm consists of two steps: 
1) creating the calibration matrix to correct the readout errors and
2) computing the corrected expectation values.
We utilized the Qiskit framework 
\cite{qiskit_error_mitigation,Qiskit_GitHub} to perform readout error mitigation. 
\par
The ZNE method estimates the result of noiseless quantum simulation from the runs at amplified noise levels.
Let $C$ be our original circuit consisting of $M$
gates: $C = U_{1}\cdots U_{M}$. We define the circuit $C^{(k)} = U_{1}^{(k)}\cdots U_{N}^{(k)}$ by replacing gates $U_i$ in $C$ with 
\begin{equation}
U_{i}^{(k)} =
    \begin{cases}
        { \ (\mathrm{CNOT})^{2k+1}\quad \mathrm{if} \quad U_{i} =\mathrm{CNOT}}\\
        { \ U_{i}  \hspace*{20mm}  \mathrm{otherwise}}
    \end{cases}.
\end{equation}
The circuit $C^{(k=0)}$ is the original circuit $C$.
Roughly the noise level of the circuit  $C^{(k>0)}$ is $\lambda:=2k+1$ times larger than that of $C^{(k=0)}$.
We define $E(\lambda=2k+1)$ as the expectation value of $Q_1^{+}$ measured with the circuit $C^{(k)}$.
We fit the unmitigated simulation results at $\lambda = 1,3\ (\Longleftrightarrow k = 0,1)$ with the linear function
$c_1 \lambda + c_0$ 
to extrapolate to $\lambda=0$ and obtain the mitigated expectation value $E(\lambda=0)=c_0$. A more general discussion can be found in \cite{ZNE}.

\par

\begin{figure}[t]
\begin{center}
\includegraphics[scale=0.78]{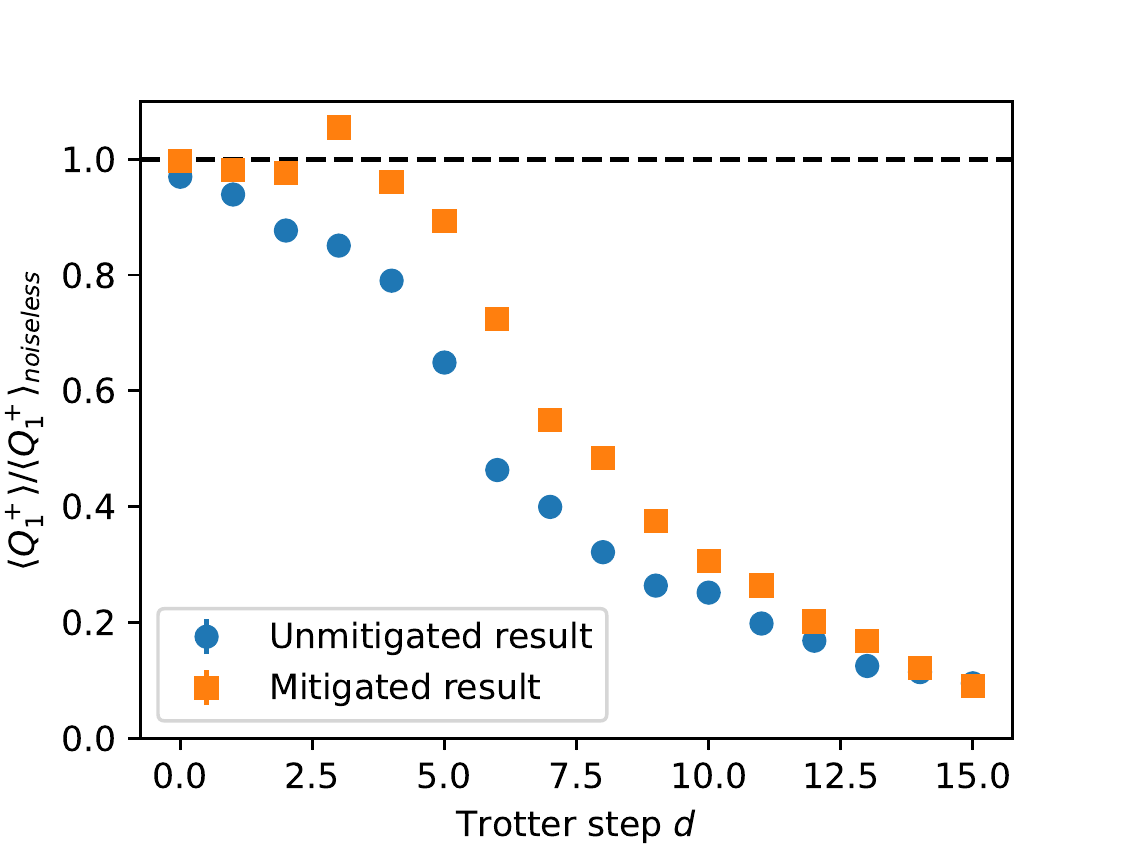}
\caption{%
Error mitigation for the expectation values of  $Q_1^+$ on 4 sites obtained by {\tt ibm\_kawasaki}. 
The mitigated results were obtained by applying readout error mitigation and the ZNE method.
Plotted results are normalized by the noiseless value $-3.8 =:\langle Q_1^{+}\rangle_{\mathrm{noiseless}}$ of $\langle Q_1^{+}\rangle$.
The data for the unmitigated and mitigated results were taken independently.
Error bars representing statistical uncertainties are hidden by markers.
}
\label{fig:EM_Kawasaki}
\end{center}
\end{figure}

We plot in FIG.~\ref{fig:EM_Kawasaki} the unmitigated and mitigated results for the time evolution of $Q_1^{+}$ with {\tt ibm\_kawasaki} on 4 sites.
The expectation values are  normalized by the noiseless value $-3.8$ of $\langle Q_1^{+}\rangle$.
We see that the results are well mitigated up to around $d=4$, after which the mitigated results approach the unmitigated ones.
This is in accord with the general fact that error mitigation is impossible when the effect of noise has accumulated~\cite{2022npjQI...8..114T,2022arXiv220809178T,2022arXiv220403455D,2021arXiv210901051W,2022arXiv220809385T,2022arXiv221011505Q}.
With devices with lower noise rates, we may use the error-mitigated expectation values of conserved charges as benchmarks for the mitigation algorithms, generalizing the discussion in section~\ref{sec:benchmarking}.


\bibliographystyle{utphys}
\bibliography{refs}

\end{document}